\documentclass[3p,number]{elsarticle}

\usepackage[cmex10]{amsmath}
\usepackage{graphicx} 
\usepackage{color} 
\usepackage{multirow}
\usepackage{placeins}
\usepackage[hyphens]{url}

\newcommand{\pp}{\widetilde{p}}

\usepackage[normalem]{ulem}
\usepackage{ifthen}
\newboolean{trackchanges}
\setboolean{trackchanges}{false}
\newboolean{revisedmanuscript}
\setboolean{revisedmanuscript}{true}
\ifthenelse{\boolean{trackchanges}}
{
	\ifthenelse{\boolean{revisedmanuscript}}
	{
	\newcommand{\inserta}[1]{\textcolor{blue}{#1}}
	\newcommand{\sustituye}[2]{\textcolor{blue}{#2}}
	\newcommand{\elimina}[1]{}
	}
	{	
	\newcommand{\inserta}[1]{\textcolor{blue}{#1}}
	\newcommand{\sustituye}[2]{\textcolor{red}{\sout{#1}}\textcolor{blue}{#2}}
	\newcommand{\elimina}[1]{\textcolor{red}{\sout{#1}}}
	}
}
{
	\newcommand{\inserta}[1]{{#1}}
	\newcommand{\sustituye}[2]{{#2}}
	\newcommand{\elimina}[1]{}
}

\begin{document}

\title{Economic feasibility of virtual operators in 5G via network slicing}

\author[upv,ups]{Erwin J. Sacoto-Cabrera}
\ead{ersacab@doctor.upv.es}

\author[upv]{Luis~Guijarro\corref{cor1}}
\ead{lguijar@dcom.upv.es}

\author[upv]{Jose~R.~Vidal}
\ead{jrvidal@upv.es}

\author[upv]{Vicent~Pla}
\ead{vpla@upv.es}

\cortext[cor1]{Corresponding author}

\address[upv]{Universitat Polit\`ecnica de Val\`encia, Spain}

\address[ups]{GIHP4C, Universidad Polit\'ecnica Salesiana-Sede Cuenca, Ecuador}

\begin{abstract}
The provision of services by more than one operator over a common network infrastructure, as enabled by 5G network slicing, is analyzed. 
Two business models to be implemented by a network operator, who owns the network, and a virtual operator, who does not, are proposed. In one business model, named \emph{strategic}, the network operator provides service to its user base and the virtual operator provides service to its user base and pays a per-subscriber fee to the network operator. In the other business model, named \emph{monopolistic}, the network operator provides service to both user bases.

The two proposals are analyzed by means of a model that captures both system and economic features. As regards the systems features, the slicing of the network is modeled by means of a Discriminatory Processor Sharing queue. As regards the economic features, the incentives are modeled by means of the user utilities and the operators' revenues; and game theory is used to model the strategic interaction between the  users' subscription decision and the  operators' pricing decision.
In both business models, it is shown that the network operator can be provided with the appropriate economic incentives so that it acquiesces in serving the virtual operator's user base (monopolistic model) and in allowing the virtual operator to provide service over the network operator's infrastructure (strategic model). From the point of view of the users, the strategic model results in a higher subscription rate than the monopolistic model.
\end{abstract}


\maketitle

%--------------------------------------------------------------------------
%--------------------------------------------------------------------------
\section{Introduction}\label{sec:intro}
%--------------------------------------------------------------------------
%--------------------------------------------------------------------------

% Contexto

The current mobile network architecture uses a relatively monolithic access and transport framework to accommodate a variety of services such as mobile traffic for smart phones, Over-The-Top content, feature phones, data cards, and embedded Machine-to-Machine devices. It is anticipated that this architecture will not be flexible and scalable enough to support the coming 5G network, which demands very diverse use cases and sometimes extreme requirements in terms of performance, scalability and availability. Furthermore, the introduction of new network services should me made more efficiently~\cite{ngmn2015}.
In the above scenario, network slicing is gaining an increasing importance as an effective way to introduce flexibility in the management of network resources. A network slice is a collection of network resources, selected in order to satisfy the requirements of the service(s) to be provided by the slice.

Among the different use cases that network slicing will enable in the coming 5G, the focus of this work is in the following: network slicing will provide an entrant operator with the mechanisms for operating a virtual network over the network infrastructure owned by an incumbent operator.

% Objetivo

In this paper, the provision of services by more than one operator over a common network infrastructure, as enabled by 5G network slicing, is analyzed. More specifically, one of the operators owns the infrastructure (hereafter, the Network Operator) and the other one does not (hereafter, the Virtual Operator). Our hypothesis is that the Network Operator can be provided with the appropriate economic incentives so that it acquiesces in allowing the Virtual Operator to provide service over the Network Operator's infrastructure.

To test the above hypothesis, two proposals are made for the business models to be implemented by the Network Operator and the Virtual Operator, and the proposals are analyzed by means of an economic model. This model captures the user utilities, the operators' revenues and the strategic interaction between the  users' subscription decision and the  operators' pricing decision.

% Contribuciones

The main contributions of this paper are the following.
\begin{enumerate}
\item The provision of service by two operators over a common infrastructure is modeled by means of a Discriminatory Processor Sharing (DPS) queue, two business models for the operators are proposed and the strategic interaction between the users and each operator is analyzed by means of Game Theory.
\item The equilibrium outcome in each business model is analyzed as a function of the most relevant parameters, which are the user sensitivity to the delay, the user-base service priority and the per-user fee paid by the Virtual Operator.
\item The conditions are established under which the business models are feasible.
\end{enumerate}

% Estructura

The paper is structured as follows. A summary of the related work is provided below. Section~\ref{sec:model} describes the proposed business models and specifies the economic model that will be analyzed by means of Game Theory. Section~\ref{sec:analysis} performs the game-theory based analysis, which yields the equilibrium outcome of the strategic interaction. Section~\ref{sec:results} discusses the effect of the different parameters on the equilibrium outcome. And Section~\ref{sec:conclusions} draws the main conclusions and points out some future research issues.

%--------------------------------------------------------------------------
\subsection{Related work}\label{sec:related}
%--------------------------------------------------------------------------

%\sustituye{There are several works which address the economic feasibility of the service provision over a general facility, where the facility is modeled as a queueing system. Without the aim of being exhaustive, we refer to Naor~\cite{naor1969}, one of the earliest works, the brilliant and comprehensive work of Allon and Federgruen~\cite{allon2008}, and the two books by Hassin~\cite{hassin2003,hassin2016}.}
\inserta{
There are many works which address the economic feasibility of the service provision over a general facility, where the facility is modeled as a queueing system. This research objective and this general model are the methodological framework of our work. Without claiming to be exhaustive, three notable contributions are referred here. Naor~\cite{naor1969} is one of the earliest works in analyzing the strategic behavior of customers in a queueing system. Allon and Federgruen~\cite{allon2008} provide a comprehensive analysis of price and service competition between servers under different modeling assumptions for the server capacity costs and for the customers demand. And the two books by Hassin~\cite{hassin2003,hassin2016} provide complete and updated surveys on the modeling and analysis of the rational behavior of users and servers in queueing systems.
}\label{tti:R1c2}\label{tti:R2c3}

More specifically, there are some works where the service under study is a network service and the infrastructure is then a network. Two works closely related to our work are~\cite{mandjes2003} and~\cite{hayel2005}, which analyze the provision of service to different user classes under a common infrastructure, as in our work. In~\cite{mandjes2003}, the network provides service differentiation to voice and data users, and it is modeled by a priority queue. Likewise, in~\cite{hayel2005}, the network is modeled by a DPS queue, which provides the desired flexibility for analyzing the provision of DiffServ services over the Internet. There is, however, an important difference between the analysis carried out in these two works and the analysis of our work: in~\cite{mandjes2003} and~\cite{hayel2005}, there is only one operator that owns and operates the network, so that there is no concern about the sharing of the infrastructure, while here, two operators are explicitly modeled, one of which owns the infrastructure, so that the infrastructure sharing problem can be tackled.

The authors have previously analyzed the economic feasibility of infrastructure sharing scenarios. 
In~\cite{guijarro2013}, the context was a cognitive radio networks, where a primary operator owned the infrastructure and a secondary operator was allowed to opportunistically access the infrastructure. The different access priorities that each operator's users had were modeled by a priority queue. 
In~\cite{sanchis2017}, the context was the provision of Human-Type Communication (HTC) and Machine-Type Communication (MTC) services, where the HTC operator owned the infrastructure. The infrastructure was also modeled by a priority queue. 
Finally, in~\cite{sacoto2018}, the context was the sharing of spectrum for supplying mobile communication services, where different spectrum sharing agreements were comparatively assessed.   
This work progresses beyond the three previous scenarios, and it tackles the 5G context. The flexibility required by the use cases and business models envisioned in 5G networks are captured by a more realistic and complex network model, which is based on a DPS queue.

%--------------------------------------------------------------------------
%--------------------------------------------------------------------------
\section{Model Description}\label{sec:model}
%--------------------------------------------------------------------------
%--------------------------------------------------------------------------

Two business models are proposed here for a Network Operator (NO, aka Operator~1) and a Virtual Operator (VO, aka Operator~2).  In both of them, each operator has its own subscriber base, so that there is no competition for the users. The NO operates a network infrastructure, but the VO does not. The NO's network then supports the service provision to the two user bases.

Each subscriber base generates a revenue for its operator. The number of subscribers depends on the quality of service (QoS) received and the price charged. The QoS depends on the network capacity and on how the operators share this capacity; both aspects are exogenously set. The price is, however, posted by the operator. The price is set by the operator in order to maximize its profits.

The difference between the two business models lies in the following.\label{businessmodel}
\begin{itemize}
\item In the first business model, referred to as \textit{monopolistic}, the NO provides service to both its own subscriber base and the VO's subscriber base.
\item In the second business model, referred to as \textit{strategic}, each operator provides service to its own subscriber base; additionally, the NO gets a revenue per VO's subscriber. 
\end{itemize}
The two business models will be assessed against a \textit{baseline scenario}, where the NO provides service to its own subscriber base supported by the network and no service is provided to the VO's subscriber base.

These business models are formally specified in this section in order to provide a basis for the analysis. First, the model for the service provided by the network infrastructure is described. And second, the model for the economic incentives of the users and the operators (NO and VO) is detailed.

A summary of the notation used in this paper is given in Table~\ref{T0}.

\begin{table}%[!t]
%\footnotesize
\caption{Summary of Notation}
\centering
\label{T0}
%\resizebox{0.8\textwidth}{!}{
\begin{tabular}{lccc}
\hline \hline
%\footnotesize
&  & Eq.            & Page  \\  \hline  
%\textbf{General Model}   &    &    &   \\  
Mean packet system time & $T_i$ & \eqref{T_1} & \pageref{T_1}\\
Mean packet service rate & $\mu$&  - &  \pageref{mu} \\   
Individual mean packet rate & $\lambda_d$ & - & \pageref{lambda_d}\\                        
NO's number of subscribers &$n_1$& - &\pageref{n_i} \\   
VO's number of subscribers &$n_2$ & - &\pageref{n_i} \\  
VO's priority  & $\gamma$ & - & \pageref{gamma} \\   
QoS perceived by the users & $Q_i$ &\eqref{qos} &\pageref{qos}  \\   
User sensitivity to delay    &  $\alpha_i$ & - &\pageref{alpha}  \\   
Conversion factor  &  $c$ &  - &\pageref{c}   \\
User utility & $U_i$ & \eqref{utility} &\pageref{utility}   \\                              
Price charged to Operator~$i$'s subscriber base &$p_i$ & - & \pageref{p_i}   \\ 
NO's profit in the monopolistic business model & $\Pi_m$ & \eqref{Pi_m} &\pageref{Pi_m} \\
Operator i's profit in the strategic business model& $\Pi_i$ & \eqref{Pi_1} &\pageref{Pi_1} \\  
 \hline \hline 
\end{tabular}
%}
\end{table}

%--------------------------------------------------------------------------
\subsection{System Model}\label{sec:system}
%--------------------------------------------------------------------------

\begin{figure}%[ht]
\centering
\includegraphics[width=0.8\columnwidth]{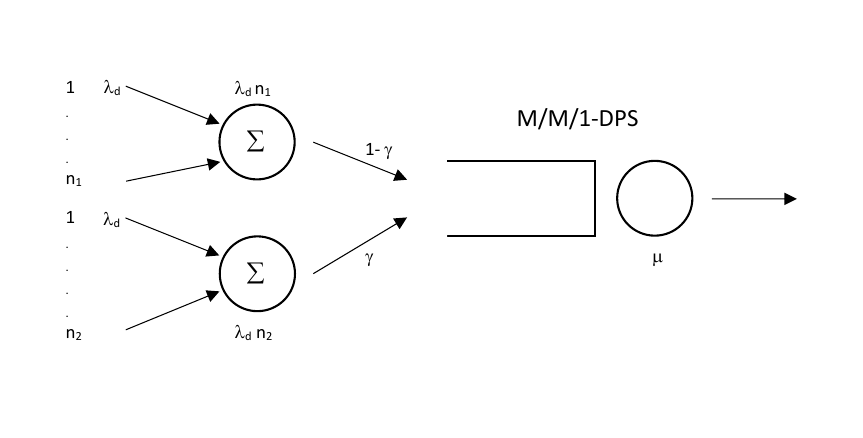}
\caption{Network Model} \label{DPSm}
\end{figure}

The network that supports the service provision to the users is modeled as one M/M/1-DPS queue, as shown in Fig.~\ref{DPSm} and explained below.

The users are modeled as independent Poisson packet sources with an individual packet generation rate $\lambda_d$\label{lambda_d}, so that each operator is offered a Poisson source of packets with rate $\lambda_i=\lambda_d \, n_i$ (where $n_i$ is the number of Operator~$i$'s subscribers\label{n_i}) and the sum of the two sources is also Poisson with rate $\lambda=\lambda_1+\lambda_2$.
The mean service times of the packets are assumed to be exponentially distributed with mean $\frac{1}{\mu}$\label{mu}. For stability reasons, it is assumed that  $\lambda<\mu$. 

\label{tti:R2c5}\inserta{To model a whole network as a single M/M/1 queue is a simplification justified by the need to obtain manageable expressions for the utility of the network users. This approach has been taken previously by~\cite{mandjes2003,hayel2004, hayel2005} in the context of the economic analysis of the internet service under the DiffServ paradigm.
Furthermore, a DPS discipline is chosen in this work in order to model the sharing of the network capacity that is enabled by network slicing. Specifically, this DPS queue manages two priorities and each subscriber base receives service for its packets at an instantaneous rate proportional to its priority. This choice allows to model the distribution of the service priority between user classes in a more flexible manner than simpler disciplines such as Priority Queueing. The modeling choice of DPS is made also in~\cite{hayel2005}, while that of Priority Queueing is made in~\cite{mandjes2003}.}
A DPS queue basically works as follows: if there are $n$ customers with priorities $x_1, x_2, \ldots, x_n$ ($x_i \leq 1$ for $i = 1, ..., n$), then customer $i$ is served at a fraction $\frac{x_i}{\sum_{j=1}^{n}x_{j}}$ of the servers capacity~\cite{hassin2003}.
In our model, the two priorities are $x_1=1-\gamma$ for the packets of NO's subscribers, and $x_2=\gamma$\label{gamma} for the packets of VO's subscribers, where $x_1+x_2=1$ and $0\leq\gamma\leq 1$.
As pointed above, DPS is a flexible mechanism for the modeling of the sharing of a common resource, thanks to parameter $\gamma$. Closed-form formulas for the mean packet service times $T_{i}$ in an M/M/1-DPS queue are given by~\cite[p. 86]{hassin2003}
\begin{equation}\label{T_1}%\label{t3}
T_1=\frac{1}{	\mu-\lambda_d n_1-\lambda_d n_2} \left(1 +\frac{\lambda_d n_2(2\gamma -1)}{\mu-(1-\gamma)\lambda_d n_1-\gamma \lambda_d n_2}\right),\\
\end{equation}

\begin{equation}\label{T_2}%\label{t4}
T_2=\frac{1}{	\mu-\lambda_d n_1-\lambda_d n_2} \left(1 -\frac{\lambda_d n_1(2\gamma -1)}{\mu-(1-\gamma)\lambda_d n_1-\gamma \lambda_d n_2}\right).\\
\end{equation}

Note that Preemptive Resume Priority Queueing, if the customers of each priority were served in a processor sharing manner, is a special case of DPS, since $\gamma=0$ gives strict priority to Operator~1's subscribers and  $\gamma=1$ gives strict priority to Operator~2's subscribers~\cite{hayel2005}. 

In the baseline scenario, where only the NO's subscriber base receives service, the queue simplifies to an M/M/1-PS queue, where the mean packet service time is
\begin{equation}\label{T_0}
 T_1= \frac{1}{\mu - n_1 \,\lambda_d}.
\end{equation}

%--------------------------------------------------------------------------
\subsection{Economic model}\label{sec:economic}
%--------------------------------------------------------------------------

The users are interested in receiving the communication service that an operator provides. Each Operator~$i$'s subscriber pays a subscription price $p_i$ for the service\label{p_i}, and receives a QoS $Q_i$.  The QoS expression is proposed to be
\begin{equation}\label{qos}
Q_i \equiv c \, T_{i}^{-\alpha_i}, \qquad i=1,2,
\end{equation}
where $c\,>0$ is a conversion factor \label{c} and $T_{i}$ is the mean packet system time, which it is referred also as delay. Parameter $\alpha_i$ denotes the sensitivity to delay of Operator~$i$'s subscribers\label{alpha}. For a given delay, a greater $\alpha_i$ translates into a worse QoS. 

The expression for the utility that a user receives is proposed to be given by the QoS in monetary units minus the price charged by the operator for the service: 
\begin{equation}\label{utility}
U_i\equiv c \, T_{i}^{-\alpha_i} -p_i, \qquad i=1,2.
\end{equation}
\inserta{It is assumed that $0 \leq \alpha_i \leq 1$, so that the utility is a concave function of the performance metrics, $1/T_i$, which is a usual assumption in telecommunication service modeling~\cite{reichl2011}.}\label{tti:R2c11b}

Finally, a user utility value equal to zero corresponds to a user who does not subscribe to the service.
\label{tti:R2c6}\inserta{The expressions for the user utility~\eqref{qos} and~\eqref{utility} has been previously adopted by~\cite{mandjes2003,hayel2004} and by the authors in~\cite{guijarro2013,sanchis2017}.}

As regards the operators, \inserta{the profits are defined as the revenues minus the costs}. Specifically, the revenues that each operator gets and the costs that each operator incurs depend on the business model, as defined in page~\pageref{businessmodel}:
\begin{description}

\item[Baseline scenario:] The NO charges a price $p_1$ to its own subscriber base. Assuming that the NO does not incurs costs, its profit is given by
\begin{equation}\label{Pi_0}
\Pi_0= n_1  \, p_1.
\end{equation}

\item[Monopolistic business model:] The NO charges a price $p_1$ to its own subscriber base, and a price $p_2$ to the VO's subscriber base. Assuming that the NO does not incurs costs, its profit is given by
\begin{equation}\label{Pi_m}%\label{pMON}
\Pi_{m}= n_1 \, p_1 + n_2  \,p_2.
\end{equation}

\item[Strategic business model:] The NO charges a price $p_1$ to its own subscriber base, and the VO charges a price $p_2$ to its own subscriber base. In addition, the VO pays a fee $\delta$ to the NO for each VO's subscriber. Otherwise, no costs are incurred by neither the NO nor the VO. The profits are then given by
\begin{equation}\label{Pi_1}%\label{pi1}
\Pi_1= n_1  \, p_1 + n_2 \,\delta
\end{equation}
\begin{equation}\label{Pi_2}%\label{pi2}
\Pi_2= n_2  \,p_2 - n_2\,\delta.
\end{equation}

\end{description}

The assumption that no costs are incurred by the operators is made for simplicity, without any loss of generality. As regards the operating costs, their inclusion would not provide additional insight since they do not depend on the service price, while it makes the expression of profits less explicit. 
As regards the investment costs, they are considered constant, given that the time scale on which the NO can adapt the network capacity $\mu$ is longer than the time scale at which prices vary~\cite{mandjes2003, sanchis2017}.

%--------------------------------------------------------------------------
\subsection{Game Model}\label{sec:game}
%--------------------------------------------------------------------------

In the three scenarios, strategic interactions can be identified between the subscription decision of the users and the pricing decisions of the operators.

\begin{description}

\item[Baseline scenario:]~\\
\begin{itemize}
\item The subscription decisions of NO's users are influenced by NO's pricing decision.

\item NO's profit depends on the subscription decisions of NO's users.
\end{itemize}

\item[Monopolistic business model:]~\\

\begin{itemize}
\item The subscription decisions of NO's users and VO's users are influenced by NO's pricing decision.

\item The subscription decisions of NO's users depend on the subscription decisions of VO's users through the factor $Q_1$. And the subscription decision of VO's users depend on the subscription decisions of NO's users through the $Q_2$ factor.

\item NO's profit depends on the subscription decisions of NO's users and of VO's users.
\end{itemize}

\item[Strategic business model:]~\\

\begin{itemize}
\item The subscription decisions of NO's users are influenced by NO's pricing decision. And the subscription decision of VO's users are influenced by VO's pricing decision.

\item The subscription decisions of NO's users depend on the subscription decision of VO's users through the factor $Q_1$. And the subscription decision of VO's users depend on the subscription decisions of NO's users through the $Q_2$ factor.

\item VO's profit depends on the subscription decision of VO's users. \inserta{And NO's profit depends on the subscription decisions of NO's users and of VO's users.}\label{tti:R2c7}

\item NO's profit is influenced by VO's pricing decision, indirectly through the subscription decision of VO's users. And vice versa.
\end{itemize}

\end{description}
These strategic interactions are amenable to analysis by means of Game Theory, where the players are the  user base and the NO in the baseline scenario; the two user bases and the NO in the monopolistic scenario; and the two user bases and the two operators in the strategic scenario. The incentives are the utilities for each user base and the profits for each operator.

The proposed game model is a two-stage game, with a different structure for each scenario:
\begin{description}
\item[Baseline scenario:] Stage I is comprised of one player (NO), which fixes $p_1$. Stage~II is comprised by its user base, within which each user chooses whether to subscribe or not.
\item[Monopolistic business model:] Stage I is comprised of one player (NO), which fixes both $p_1$ and $p_2$.
\item[Strategic business model:] Stage I is comprised of two players (NO and VO), each one fixing its service price.
\end{description}
In both the monopolistic and the strategic models, stage~II is comprised of the two user bases, within which each user chooses whether to subscribe or not to its operator.

The solution of the game is an equilibrium decision or strategy for each player. The equilibrium concept used here is the Nash equilibrium, where no player has an incentive to deviate from its equilibrium strategy, provided that the rest of the players are playing the equilibrium strategy.

This two-stage game is solved using backward induction~\cite{maschler2013}\sustituye{, which means that at Stage I players proceed anticipating the solution of Stage II.}{. Since players at stage~II choose their action with the knowledge of stage~I players choice, in the equilibrium stage~I players will anticipate the choice of stage~II players. This provides a rationale for solving the two-stage game by first solving the equilibrium of stage~II for given known stage~I player actions, and then proceeding backwards to solve the equilibrium of stage~I with the knowledge of the stage~I best response. The equilibrium computation is presented below following this ordering: first, stage~II; and second, stage~I.}\label{tti:R2c8}

%--------------------------------------------------------------------------
\subsubsection{Stage II - Users subscription}

In stage~II each user takes his/her own subscription decision, trying to maximize the utility he/she gets from either subscribing to the operator or not. 
Each user in Operator~$i$'s user base will observe price $p_i$ and will make a subscription decision based on the utility he/she would get from each alternative: $U_i$ (see~\eqref{utility}) if he/she subscribes, or 0 if he/she does not.

Assuming that the number of users is high enough, the individual subscription decision of one user from Operator~$i$'s user base does not affect the utility of the rest of Operator~$i$'s user base. 
\label{tti:R2c9}\sustituye{Under these conditions, the equilibrium reached is the one postulated by Wardrop~\cite{wardrop1952}. Basically, at a \emph{Wardrop equilibrium}, the utility that every user gets is equalized between the decision outcomes. That means that, as regards Operator~$i$'s user base, either $U_i=0$ and some users subscribe ($n_i\geq 0$), or $U_i<0$ and no user subscribes ($n_i = 0$).}{Under these conditions, the equilibrium reached is the one postulated by Wardrop~\cite{wardrop1952} in 1952 as a rule to solve the traffic assignment problem, i.e., a problem that concerns the selection of routes between origins and destinations in transportation networks. Specifically, Wardrop's first principle is the relevant one, which says that:
\emph{The journey times on all routes actually used are equal, and less than those which would be experienced by a single vehicle on any unused route.} 
Basically, at a \emph{Wardrop equilibrium}, the utility that every user gets is equalized between the alternative effectively chosen by the users.
That means that, as regards Operator~$i$'s user base, either (1) $U_i=0$ (i.e., subscription and no-subscription utilities are equal) and some users subscribe and some other users do not ($n_i^*\geq 0$), or (2) $U_i<0$ and no user subscribes ($n_i^* = 0$; i.e., no user chooses the option with less than zero utility). Note that the third alternative ($U_i>0$) can not be sustained under~\eqref{utility} and~\eqref{T_1}--\eqref{T_0}, since an increase in $n_i$ causes an increase in $T_i$ and will eventually settle in $U_i=0$. }

For the baseline scenario, only two possible cases can be identified:
\begin{itemize}
\item Case I:
\begin{equation} \label{wi1}
U_1=0, \text{ and } n_1^* \geq 0.
\end{equation}
\item Case II: 
\begin{equation} \label{wi2}
U_1<0, \text{ and } n_1^* = 0.
\end{equation}
\end{itemize}

Whereas for the monopolistic and strategic business models, four possible cases can be identified:
\begin{itemize}
\item Case I:
\begin{equation} \label{wr1}
U_1=0, \, U_2=0 \text{, and } n_1^* \geq 0, \ n_2^* \geq 0.
\end{equation}
\item Case II: 
\begin{equation} \label{wr2}
U_1=0, \, U_2<0 \text{, and } n_1^* \geq 0, \ n_2^* = 0.
\end{equation}
\item Case III: 
\begin{equation} \label{wr3}
U_1<0, \, U_2=0 \text{, and } n_1^* = 0, \ n_2^* \geq 0.
\end{equation}
\item Case IV: 
\begin{equation} \label{wr4}
U_1<0, \, U_2<0 \text{, and } n_1^* = 0, \ n_2^* = 0.
\end{equation}
\end{itemize}
As detailed in Section~\ref{sec:analysis}, the above four outcomes express $n_1^*$ and $n_2^*$ as functions of $p_1$ and $p_2$, i.e., $n_1^* (p_1, p_2)$ and $n_2^* (p_1, p_2)$.

%--------------------------------------------------------------------------
\subsubsection{Stage I - Operator/s pricing decision}\label{sec:stageI}

In the baseline scenario, the NO chooses $p_1$ in order to maximize its profits, $\Pi_0$, given by~\eqref{Pi_0}. The NO anticipates that the users subscription will set into the equilibrium described in~\eqref{wi1}--\eqref{wi2}, so that $\Pi_0$ will be a function of $p_1$. The profit maximizing price $p_1^*$ is given as
\begin{equation}\label{p_0}
p_1^*=\arg\max_{p_1}\ \Pi_0(p_1).
\end{equation}

In the monopolistic business model, the NO chooses $p_1$ and $p_2$ in order to maximize its profits,$\Pi_m$, given by~\eqref{Pi_m}. The NO anticipates that the users subscription will settle into the equilibrium described in~\eqref{wr1}--\eqref{wr4}, so that $\Pi_m$ will be a function of $p_1$ and $p_2$. The profit maximizing prices $p_1^*$ and $p_2^*$ are given as
\begin{equation}\label{p_opt}%\label{MAX}
\lbrace p_1^*, p_2^*\rbrace=\arg\max_{p_1,p_2}\ \Pi_m(p_1,p_2).
\end{equation}

In the strategic business model, each operator is aware not only of the users subscription equilibrium in stage~II and of its profit function, but also of the rational behavior of the other operator. The profit maximizing price for each operator will then depend also on the other operator's choice, i.e., it will be given by a best response function (BR):
\begin{equation}\label{BR1}
BR_1(p_2)=\arg\max_{p_1} \Pi_1(p_1,p_2),
\end{equation}
\vspace{-5mm}
\begin{equation}\label{BR2}
BR_2(p_1)=\arg\max_{p_2} \Pi_2(p_1,p_2).
\end{equation}

The Nash equilibrium at stage~I will be a pair of prices $p_1^*$ and $p_2^*$ such that each operator is fixing a best response price to the other operator's price anticipating stage~II equilibrium, i.e., the solution of the following system of equations:
\begin{equation}\label{NE1}%\label{42}
p_1^*=BR_1(p_2^*),
\end{equation}
\vspace{-5mm}
\begin{equation}\label{NE2}%\label{43}
p_2^*=BR_2(p_1^*).
\end{equation}

%--------------------------------------------------------------------------
%--------------------------------------------------------------------------
\section{Analysis}\label{sec:analysis}%\label{ANALI}
%--------------------------------------------------------------------------
%--------------------------------------------------------------------------

In this section, first, the Wardrop equilibrium for stage~II for the three scenarios is obtained analytically. Then, the solution of stage~I for the baseline scenario is presented. \elimina{For the monopolistic and strategic business models, the solution of stage~I is not feasible. The solution of stage~I for these two models will be performed numerically in Section~\ref{sec:results}, following the maximization problems specified in Section~\ref{sec:stageI}.}

%--------------------------------------------------------------------------
\subsection{Wardrop equilibrium: baseline scenario}
%--------------------------------------------------------------------------

The utility of a user can be obtained replacing~\eqref{T_0} in~\eqref{utility}, yielding
\begin{equation}\label{U0}
U_1= c\cdot (\mu -n_1 \cdot \lambda_d)^{\alpha_1}-p_1.
\end{equation}
With this expression for the user's utility, the Wardrop equilibrium expressed in~\eqref{wi1} and~\eqref{wi2} results in the following specifications:
\begin{equation}\label{n1Ibase}
n_1^*=
\begin{cases}
\frac{\mu -\left( \frac{p_1}{c} \right)^{\frac{1}{\alpha _1}}}{\lambda _d} & \text{if } p_1 \leq  c \mu^{\alpha_1},  \\
0 & \text{if } p_1 >  c \mu^{\alpha_1}. 
\end{cases}
\end{equation}%\label{n1IIbase}

%--------------------------------------------------------------------------
\subsection{Wardrop equilibrium: monopolistic and strategic business models}
%--------------------------------------------------------------------------

The utility of a user belonging to each user base can be obtained replacing~\eqref{T_1} and~\eqref{T_2}  in~\eqref{utility}, yielding
\begin{equation}\label{U1}%\label{5a}
U_1= c \left(\frac{\mu -(1-\gamma) (\lambda_d n_1+ \lambda_d n_2)}{\left(\mu -\lambda_d n_1-\lambda _d n_2\right) \left(\mu  - (1-\gamma)\lambda_d n_1 - \gamma \lambda_d n_2 \right)}\right)^{-\alpha_1 } - p_1,
\end{equation}
\begin{equation}\label{U2}%\label{6a}
U_2= c \left(\frac{\mu -\gamma  \left(\lambda_d n_1+\lambda_d n_2\right)}{\left(\mu -\lambda_d n_1-\lambda_d n_2\right) \left(\mu  - (1-\gamma)\lambda_d n_1 - \gamma \lambda_d n_2 \right)}\right)^{-\alpha_2 } - p_2.
\end{equation}

With these expressions for the user's utility, the Wardrop equilibrium expressed in~\eqref{wr1}, \eqref{wr2}, \eqref{wr3} and~ \eqref{wr4} results in the following specifications:

\begin{itemize}

\item Case I : From~\eqref{wr1}, \eqref{U1} and~\eqref{U2},
\begin{equation}\label{n1Iwr1}
n_1^*=\frac{1}{\lambda_d} \left[\frac{1}{\gamma\pp_2^{-\frac{1}{\alpha_2}}-(1-\gamma)\pp_1^{-\frac{1}{\alpha_1}}}-\frac{\mu \pp_2^{-\frac{1}{\alpha_2}}}{\gamma \pp_1^{-\frac{1}{\alpha_1}} - (1-\gamma) \pp_2^{-\frac{1}{\alpha_2}}}\right]
\end{equation}
\begin{equation}\label{n2Iwr1}
n_2^*=\frac{1}{\lambda_d} \left[\frac{\mu \pp_1^{-\frac{1}{\alpha_1}}}{\gamma \pp_1^{-\frac{1}{\alpha_1}} - (1-\gamma) \pp_2^{-\frac{1}{\alpha_2}}}-\frac{1}{\gamma\pp_2^{-\frac{1}{\alpha_2}}-(1-\gamma)\pp_1^{-\frac{1}{\alpha_1}}}\right],
\end{equation}
\begin{equation}
p_1 \leq  \widehat{p}_1(p_2),
\end{equation}
\begin{equation}
p_2 \leq  \widehat{p}_2(p_1).
\end{equation}
where 
\begin{equation}
\pp_i \equiv \frac{p_i}{c}, \quad i=1,2,
\end{equation}
\begin{equation}
\widehat{p}_2(p_1) \equiv c \left[ \frac{(1-\gamma) \pp_1^{\frac{1}{\alpha_1}} + \gamma \mu }{(1-\gamma) \mu \pp_1^{-\frac{1}{\alpha_1}}+\gamma}\right]^{\alpha_2},
\end{equation}
\begin{equation}
\widehat{p}_1(p_2) \equiv c \left[ \frac{\gamma \pp_2^{\frac{1}{\alpha_2}} + (1-\gamma) \mu }{\gamma \mu \pp_2^{-\frac{1}{\alpha_2}}+(1-\gamma)}\right]^{\alpha_1}.
\end{equation}

\item Case II: From~\eqref{wr2}, \eqref{U1} and~\eqref{U2},
\begin{equation}
n_1^*=\frac{\mu -\pp_1^{\frac{1}{\alpha _1}}}{\lambda _d},
\end{equation}
\begin{equation}
n_2^*=0,
\end{equation}
\begin{equation}
p_1 \leq c \mu^{\alpha_1},
\end{equation}
\begin{equation}
p_2 >  \widehat{p}_2(p_1).
\end{equation}

\item Case III : From~\eqref{wr3}, \eqref{U1} and~\eqref{U2},
\begin{equation}
n_1^*=0,
\end{equation}
\begin{equation}
n_2^*=\frac{\mu -\pp_2^{\frac{1}{\alpha _2}}}{\lambda _d},
\end{equation}
\begin{equation}
 p_2 \leq  c \mu^{\alpha_2},
\end{equation}
\begin{equation}
p_1 >  \widehat{p}_1(p_2).
\end{equation}

\item Case IV : Finally, from~\eqref{wr4}, \eqref{U1} and~\eqref{U2}
\begin{equation}
n_1^*=0  ,
\end{equation}
\begin{equation}
n_2^*=0 ,
\end{equation}
\begin{equation}
p_1 > c \mu^{\alpha_1},
\end{equation}
\begin{equation}
p_2 > c \mu^{\alpha_2}. 
\end{equation}

\end{itemize}

Table~\ref{T4} summarizes the above expressions for the Wardrop equilibrium.

{
\renewcommand{\arraystretch}{3}
\begin{table}[!t]
%\footnotesize
\caption{Users' Subscription Wardrop equilibrium}
\centering
\label{T4}
%\resizebox{\textwidth}{!}
%{
\begin{tabular}{ccccc}
\hline \hline
%\footnotesize
\textbf{Case} 
& $n_1^*$ 
& $n_2^*$ 
& $p_1$                
& $p_2$
\\ 

\hline

I                
&  \eqref{n1Iwr1}
& \eqref{n2Iwr1}
& $ p_1\leq \widehat{p}_1 (p_2)$ & $p_2\leq \widehat{p}_2(p_1)$ 
\\

\hline

II  
& $\frac{\mu -\pp_1^{\frac{1}{\alpha _1}}}{\lambda _d}$  
& 0 
& $ p_1 \leq c \mu^{\alpha_1}$ 
& $ p_2> \widehat{p}_2(p_1) $ 
\\  
 
\hline

III 
& 0 
& $\frac{\mu -\pp_2^{\frac{1}{\alpha _2}}}{\lambda _d}$ 
& $p_1>\widehat{p}_1(p_2)$ 
& $p_2 \leq c \mu^{\alpha_2}$
\\

\hline

IV  
& 0           
& 0  
& $p_1>c \mu^{\alpha_1}$                     
& $p_2>c \mu^{\alpha_2}$    
\\ 

\hline \hline

\end{tabular}

\end{table}
}

\begin{figure}%[h]
\centering
{\includegraphics[width=0.5\columnwidth]{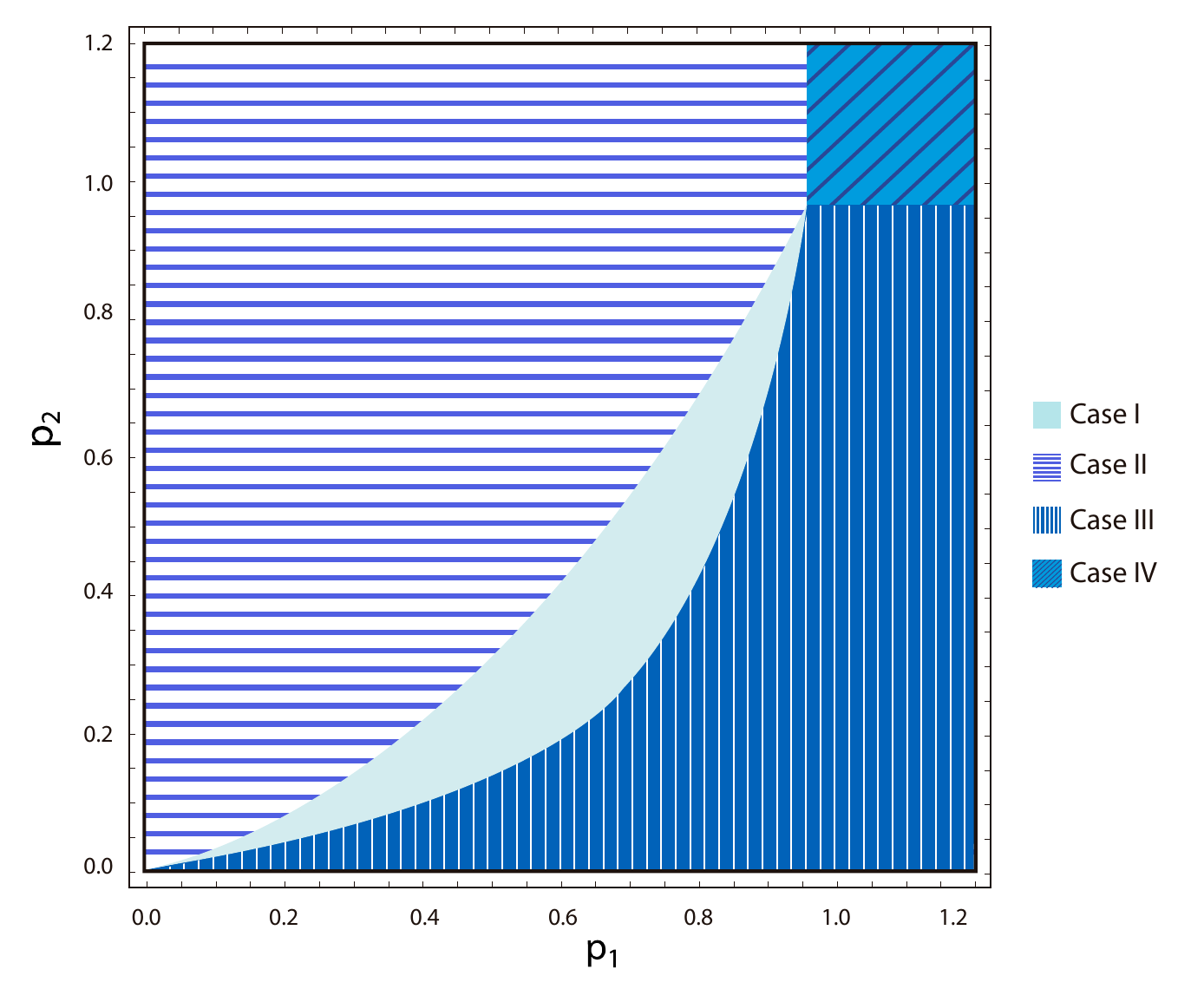}}
\caption{Wardrop equilibrium cases/regions for $\gamma=1/10$ and $\alpha_1=\alpha_2=0.8$}\label{regions1}
\end{figure}

\begin{figure}%[h]
\centering
\includegraphics[width=0.5\columnwidth]{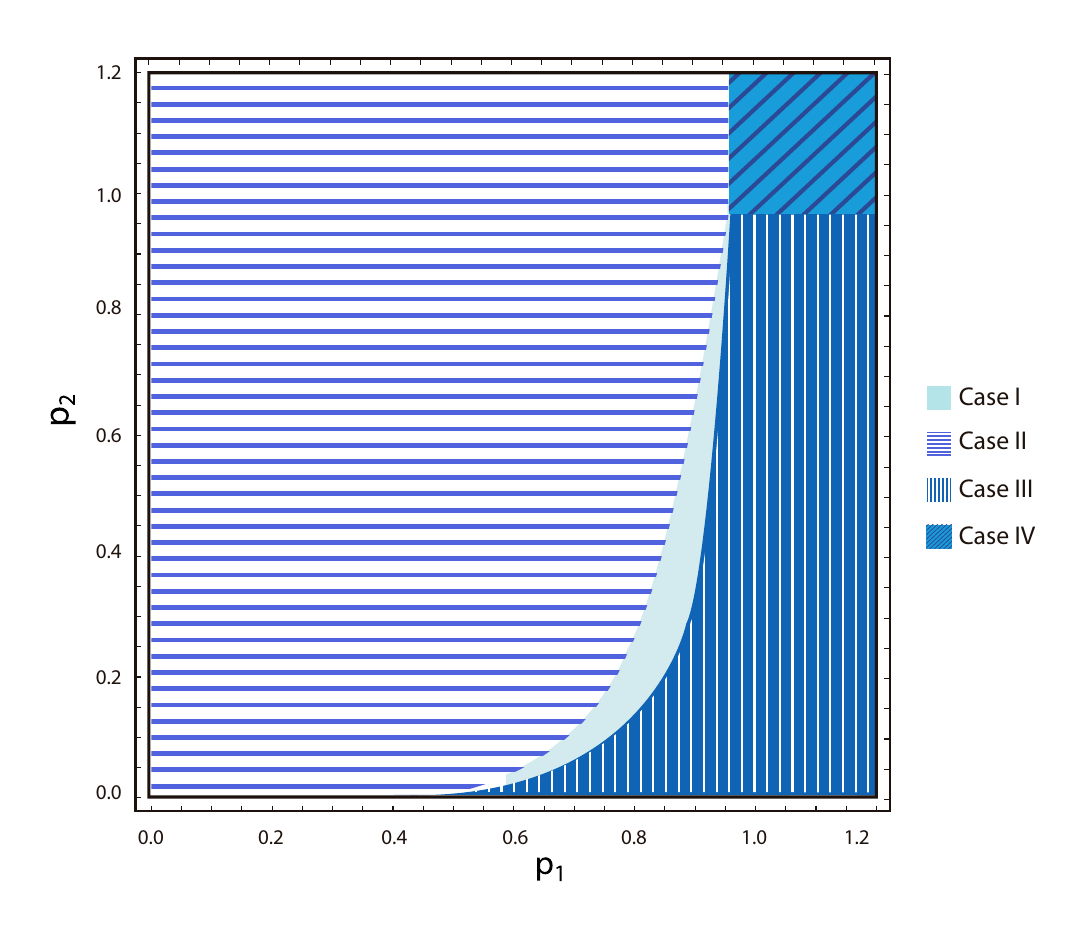}
\caption{Wardrop equilibrium cases/regions for $\gamma=1/10$ $\alpha_1=0.2$ and $\alpha_2=0.8$}\label{regions2}
\end{figure}

\begin{figure}%[!t]
\centering
\includegraphics[width=0.5\columnwidth]{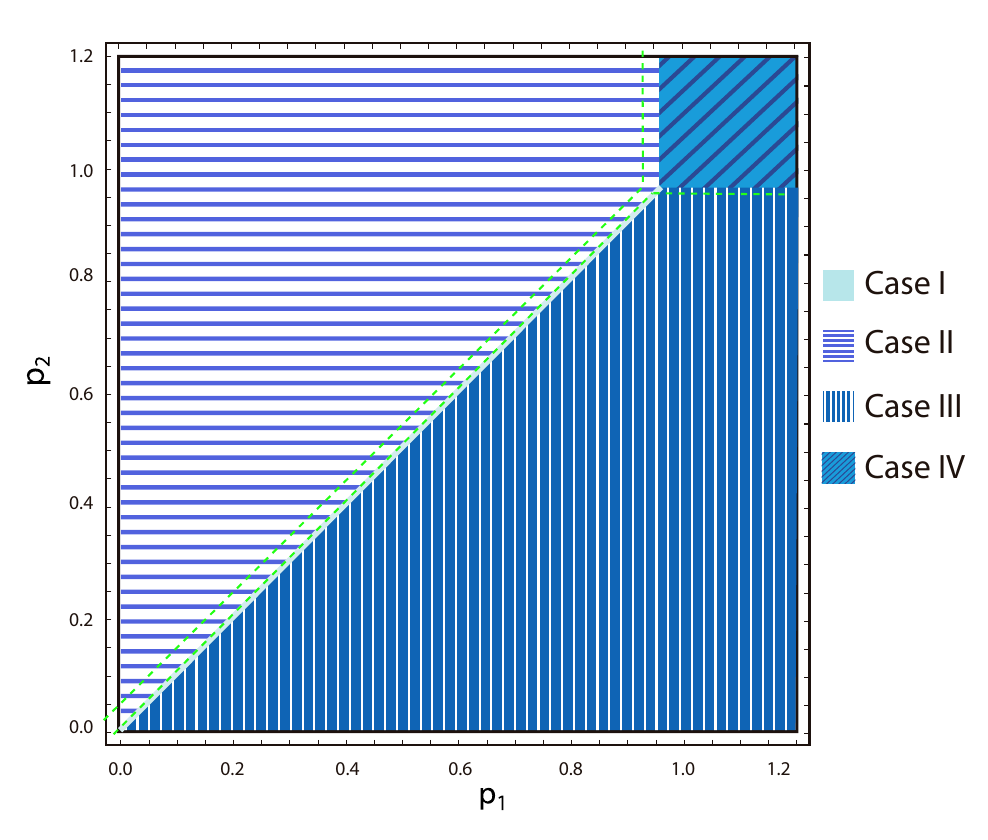}
\caption{Wardrop equilibrium cases/regions for $\gamma=1/2$ and $\alpha_1=\alpha_2=0.8$}\label{regions3}
\end{figure}

\begin{figure}%[!t]
\centering
\includegraphics[width=0.5\columnwidth]{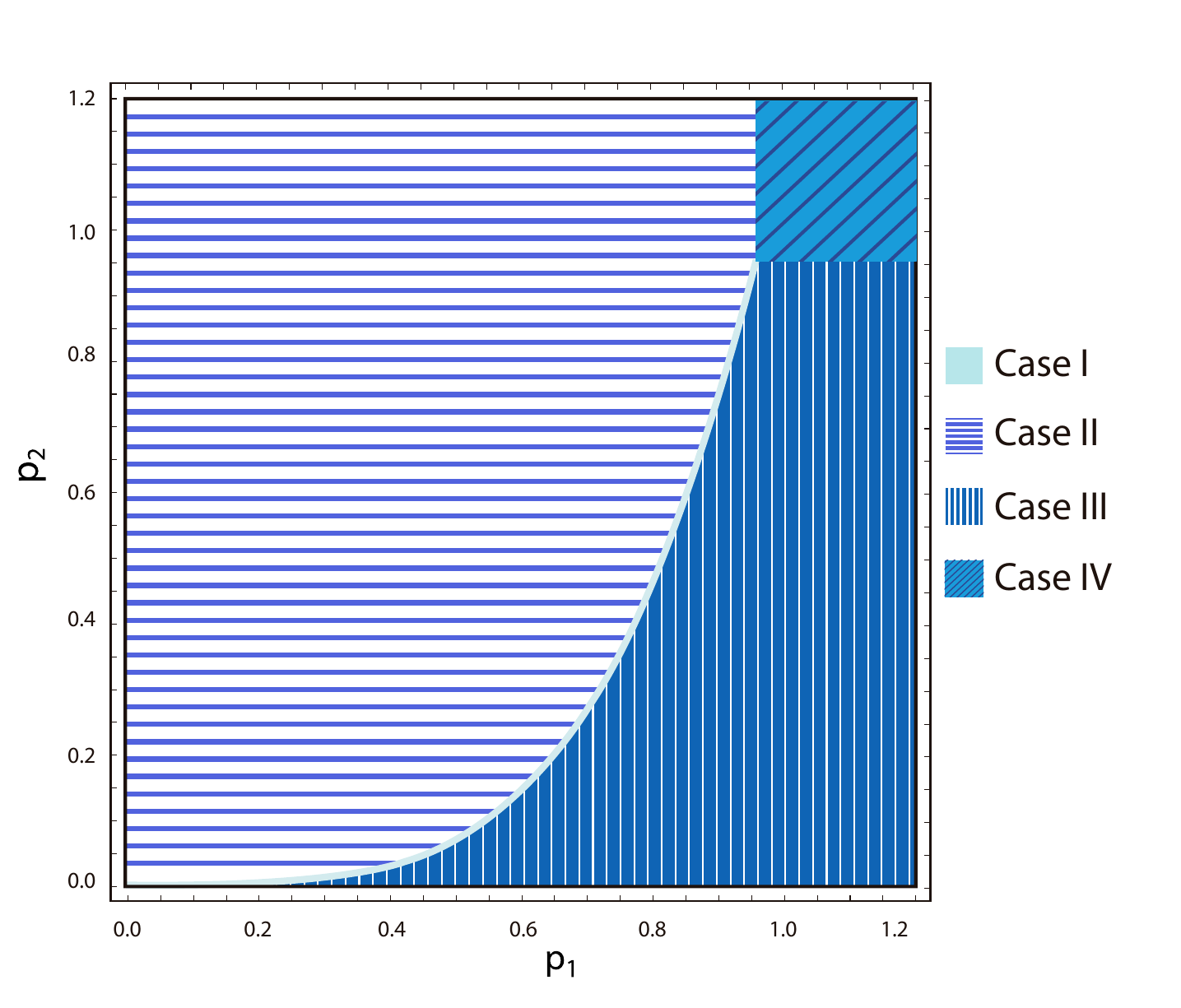}
\caption{Wardrop equilibrium cases/regions for $\gamma=1/2$, $\alpha_1=0.2$ and $\alpha_2=0.8$}\label{regions4}
\end{figure}

Figs.~\ref{regions1}, \ref{regions2}, \ref{regions3} and~\ref{regions4} show a graphical representation of the equilibrium cases in the $(p_2, p_1)$ plane for \sustituye{the following values of the parameters: $c=1$, $\mu=1$ packets/s and $\lambda_d=0.5$ packets/s and different combinations of $\alpha_1$, $\alpha_2$ and $\gamma$.}{different combinations of $\alpha_1$, $\alpha_2$ and $\gamma$. In all figures, $c=1$ and $\mu=1$ packets/s, so that Case~IV always corresponds to $p_1>1$ and $p_2>1$}. 

\label{tti:R1c3}\label{tti:R2c10}\inserta{In all figures, for a given $p_2$, a low $p_1$ drives $n_2^*$ to zero (Case~II); increasing $p_1$ eventually allows $n_2^*$ to increase, while $n_1^*$ decreases until $n_1^*$ reaches zero (Case~III). And vice versa for a given $p_1$ and $p_2$ increasing from a low value. The prices that cause the transition from Case~II through Case~I to Case~III depend on parameters $\alpha_1$, $\alpha_2$ and $\gamma$.
For a given $\alpha_2$, the higher $\alpha_1$, the lower $p_1$ that causes the transition to Case~III (see, e.g., Fig.~\ref{regions1} vs. Fig.~\ref{regions2}). In other words, the more sensitive the users, the lower the minimum price that makes them refuse to subscribe.
The influence of $\gamma$ is on the ``width'' of Case~I. The nearer $\gamma$ is to $1/2$, the thinner is the region for Case~I (see, e.g., Fig.~\ref{regions3} vs. Fig.~\ref{regions4}). Or, in other words, the more asymmetrical is the priority distribution, the wider the range of price values that result in non-zero subscriptions for both operators.}

\inserta{The equilibrium Cases described above will be hereafter referred to as equilibrium Regions.}\label{tti:R2c13}
    
%--------------------------------------------------------------------------
\subsection{Stage I analysis}
%--------------------------------------------------------------------------

For the baseline scenario, according to~\eqref{Pi_0} and given the Wardrop equilibrium for $n_1^*$ obtained in~\eqref{n1Ibase}% and in~\eqref{n1IIbase}
, the profit expression to be maximized by the NO is the following:
\begin{equation}
\Pi_0 (p_1) = 
	\begin{cases}
	\frac{1}{\lambda _d} \left[ \mu- \left( \frac{p_1}{c} \right) ^{\frac{1}{\alpha _1}} \right] p_1, & \text{if } p_1 \leq c \mu^\alpha_1,\\
	0 & \text{if } p_1 > c \mu^\alpha_1.
	\end{cases}	
\end{equation}

The optimal price that solves~\eqref{p_0} and the maximum profits are
\begin{align}
p_1^* &= c \left(  \frac{\alpha_1}{1+\alpha_1} \mu \right)^{\alpha_1},\\
\Pi_0^* = \Pi_0 (p_1^*) & = \frac{c}{\alpha_1 \lambda_d} \left(  \frac{\alpha_1}{1+\alpha_1} \mu \right)^{1+\alpha_1}.
\end{align}

\label{tti:R2c12}\inserta{For the monopolistic and strategic business models, the solution of stage~I is not analytically tractable. The solution of stage~I for these two models will be performed numerically in Section~\ref{sec:results}, following the maximization problems specified in Section~\ref{sec:stageI}.}

\FloatBarrier

%--------------------------------------------------------------------------
%--------------------------------------------------------------------------
\section{Results and discussion}\label{sec:results}
%--------------------------------------------------------------------------
%--------------------------------------------------------------------------

In this section, numerical values are computed for the prices, number of subscribers and operators' profits in the monopolistic (Section~\ref{sec:results_monopolistic}) and strategic business models (Section~\ref{sec:results_strategic}). For each business model, two cases are analyzed: the case where both user bases have the same sensitivity to the delay ($\alpha_1=\alpha_2$) and the case where they have different sensitivities ($\alpha_1\neq\alpha_2$). Finally, in Section~\ref{sec:results_feasibility}, the feasibility of each business model is analyzed, that is, which incentives each business model provides to the NO and the VO so that they are better off than in the baseline scenario.

\inserta{If not stated otherwise, the following parameter values are used:
\begin{itemize}
\item $c=1$, which is a normalizing constant for the prices and profits.
\item $\mu=1$ packet/s.
\item $\lambda_d=0.01$ packet/s, i.e., two orders of magnitude lower than $\mu$, so that the assumption that the number of users is high, which has been made in the Wardrop equilibrium, can be justified.
\end{itemize}
}\label{tti:R2c11a}

%--------------------------------------------------------------------------
\subsection{Monopolistic business model}\label{sec:results_monopolistic}
%--------------------------------------------------------------------------

The equilibrium in the monopolistic scenario, where the NO operates the network and provides service to its own subscriber base and to the VO's subscriber base, is presented and discussed. The NO's subscriber base is assigned a priority $1-\gamma$ in the use of the network capacity, while the VO's subscriber base is assigned a priority $\gamma$. 

First, the case where both user bases have the same sensitivity to the delay ($\alpha_1=\alpha_2$) is addressed, and second, the case where they have different sensitivities ($\alpha_1\neq\alpha_2$). 

%--------------------------------------------------------------------------
\subsubsection{User bases with the same sensitivity}

Parameters are set $\alpha_1=\alpha_2=\alpha=\{0.2, 0.4, 0.6, 0.8, 1\}$ and the effect of the priority $\gamma$ on the equilibrium is analyzed.

\paragraph{Prices} 

\begin{figure}%[!h]
\centering
\includegraphics[width=0.7\columnwidth]{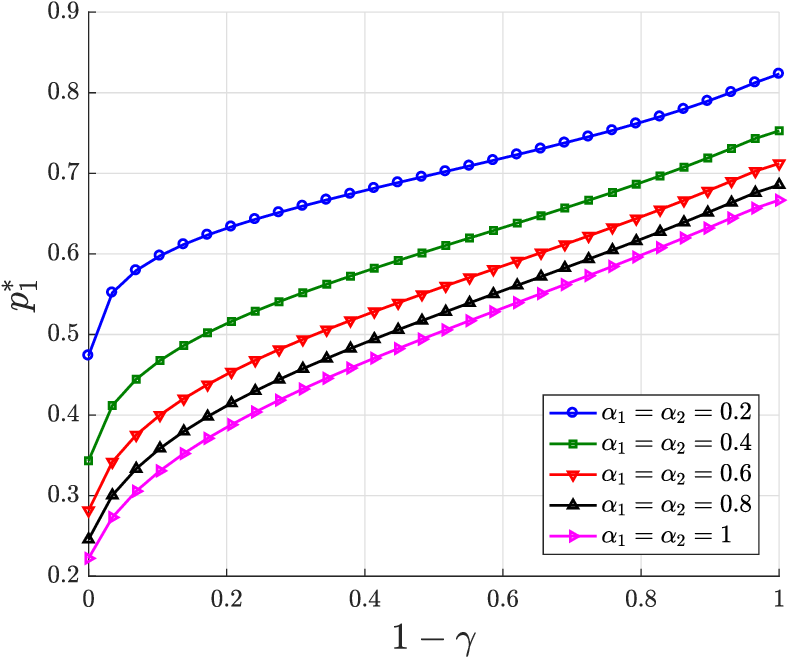}
\caption{Price $p_{1m}^*$ as a function of $1-\gamma$ for different values of the common $\alpha$ (monopolistic)} \label{p1monhom}%\label{pm1}
\vspace{-3mm}
\end{figure}

Figure~\ref{p1monhom} shows the price $p_{1m}^*$ set by the NO in the equilibrium to each subscriber base as a function of $1-\gamma$ (NO's subscriber base priority) for different values of $\alpha$. Due to symmetry, this graph also represents $p_{2m}^*$ as a function of $\gamma$ (VO's subscriber base priority).

It is seen that the service is priced higher as the service priority increases. And that the subscriber base that receives a service supported by a greater priority is priced higher (note that if $1-\gamma \leq 1/2$ then $\gamma \geq 1/2$). Consistently with this observation, when $\gamma=1/2$, both subscriber bases are priced equally. Finally, the greater the sensitivity $\alpha$, the lower the price set by the NO to the subscribers. 

\paragraph{Number of subscribers}

\begin{figure}%[!h]
\centering
\includegraphics[width=0.7\columnwidth]{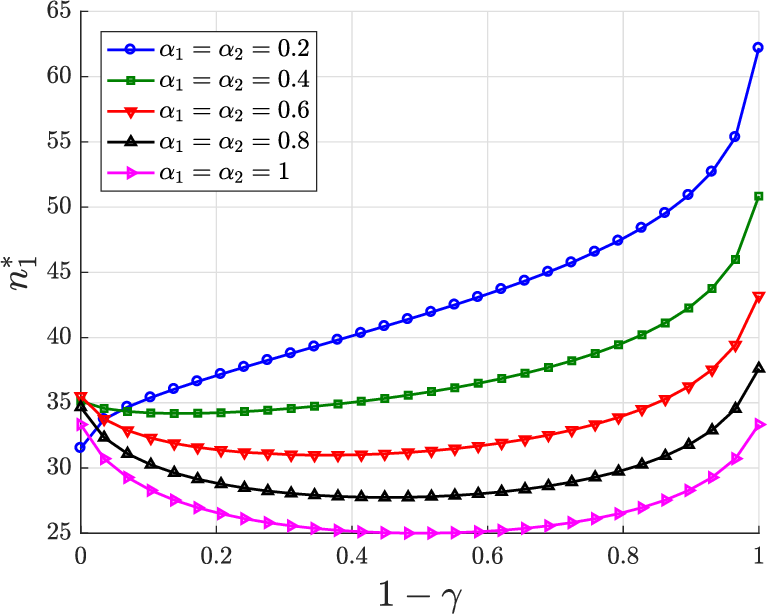}
\caption{Number of subscribers $n_{1m}^*$ as a function of $1-\gamma$ for different values of the common $\alpha$ (monopolistic)} \label{n1monhom}%\label{nm1}
\vspace{-3mm}
\end{figure}

Figures~\ref{n1monhom} shows the number of NO's subscribers, $n_{1m}^*$, in the equilibrium as a function of $1-\gamma$ (NO's subscriber base priority) for different values of $\alpha$. Due to symmetry, this graph also represents $n_{2m}^*$ as a function of $\gamma$ (VO's subscriber base priority).
The number of subscribers is a measure of the aggregate benefit that the users get, bearing in mind that the users equilibrium is at Region~I, where $U_1=U_2=0$ (see Figs.~\ref{regions1} and ~\ref{regions2}).

Two different evolutions of $n_{1m}^*$ with $1-\gamma$ can be distinguished, which depend on $\alpha$. For low sensitivity users (i.e., low values of $\alpha$, such as 0.2), the number of subscribers $n_{1m}^*$ increases as their priority ($1-\gamma$) increases. However, for high sensitivity users (i.e., $\alpha \geq 0.4$), the number of subscribers $n_{1m}^*$ first decreases, reaches a minimum and then increases; the position of this minimum depends on $\alpha$: the higher $\alpha$, the higher the priority $1-\gamma$ that yields the minimum. 
From the observation of how the price and the number of subscribers varies with the service priority, it is inferred that for both low sensitivity and high sensitivity users, a high priority translates into a sufficiently high QoS that compensates for the high price set; and that, only for high sensitivity users, a low priority also translates into a sufficiently low price that compensates for the low QoS received.

As regards the sensitivity, it is seen that, for high sensitivity users, the greater the sensitivity $\alpha$, the lower the number of subscribers. 
From the observation of how the price and the number of subscribers varies with the sensibility, it is inferred that, for high sensitivity users, higher sensitivity translates into a lower QoS that is not compensated by the lower prices. This can also be applied to low sensitivity users, except when the priority is very low.  

\begin{figure}%[!h]
\centering
\includegraphics[width=0.7\columnwidth]{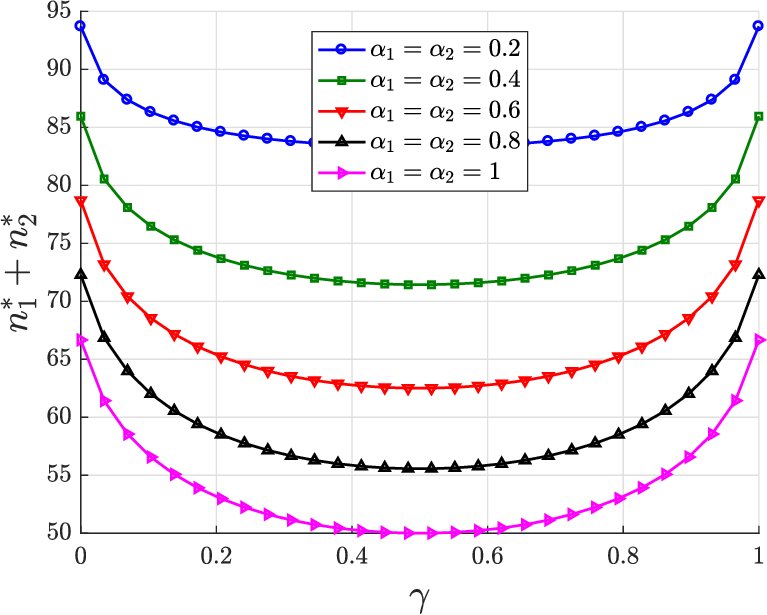}
\caption{Total number of subscribers as a function of $\gamma$ for different values of the common $\alpha$ (monopolistic)} \label{n1n2monhom}%\label{nm2}
\vspace{-3mm}
\end{figure}

Finally, Fig.~\ref{n1n2monhom} shows the total number of subscribers, i.e., $n_{1m}^*+n_{2m}^*$. Maximum subscription rates are achieved either when the service priority is equal to 1 or 0.

The conclusion is that, when the two user bases have the same sensitivity, the most favorable network configuration for the users is one where full priority is given to one user base. 

\paragraph{Operator's profit} 

\begin{figure}%[!h]
\centering
\includegraphics[width=0.7\columnwidth]{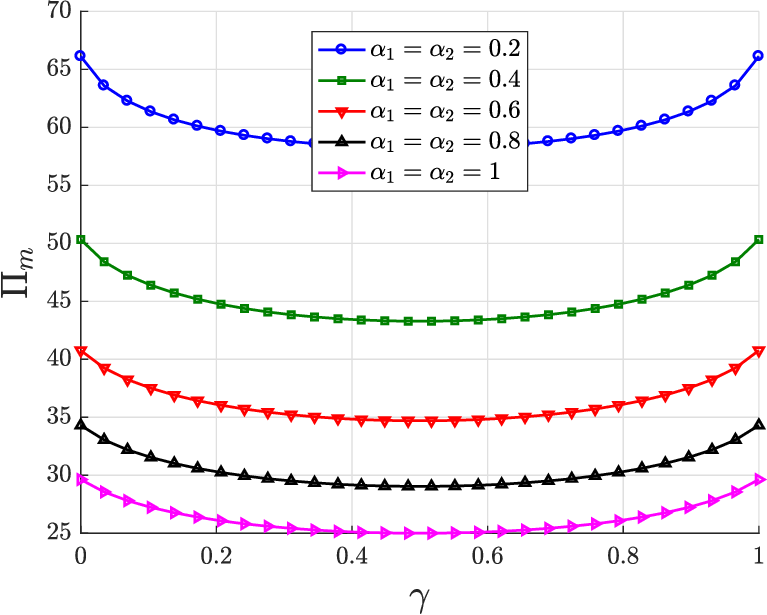}
\caption{NO's profit $\Pi_m^*$ as a function of $\gamma$ for different values of the common $\alpha$ (monopolistic)} \label{pimonhom}
\vspace{-3mm}
\end{figure}

Figure~\ref{pimonhom} shows the NO's profit as a function of $\gamma$ (VO's subscriber base priority) for different values of $\alpha$.

It is seen that the higher is the difference between the priorities of the two user bases($\vert (1-\gamma) - \gamma \vert = 2 \vert 1/2 - \gamma\vert$), the higher is the NO's profit. This is consistent with the fact that a high $1-\gamma$ gives high $p_{1m}^*$ and high $n_{1m}^*$, while a high $\gamma$ gives high $p_{2m}^*$ and high $n_{2m}^*$. 

The conclusion is that, when the two user bases have the same sensitivity, the most favorable network configuration for the operator is one where full priority is given to one user base. This agrees with the most favorable configuration from the point of view of the users.

\FloatBarrier

%--------------------------------------------------------------------------
\subsubsection{User bases with different sensitivities}

Parameters are set $\alpha_1=0.6$ and $\alpha_2=\{0.2, 0.4, 0.6, 0.8, 1\}$ and the effect of the priority $\gamma$ on the equilibrium is analyzed.

\paragraph{Prices}

\begin{figure}%[!h]
\centering
	\includegraphics[width=0.7\columnwidth]{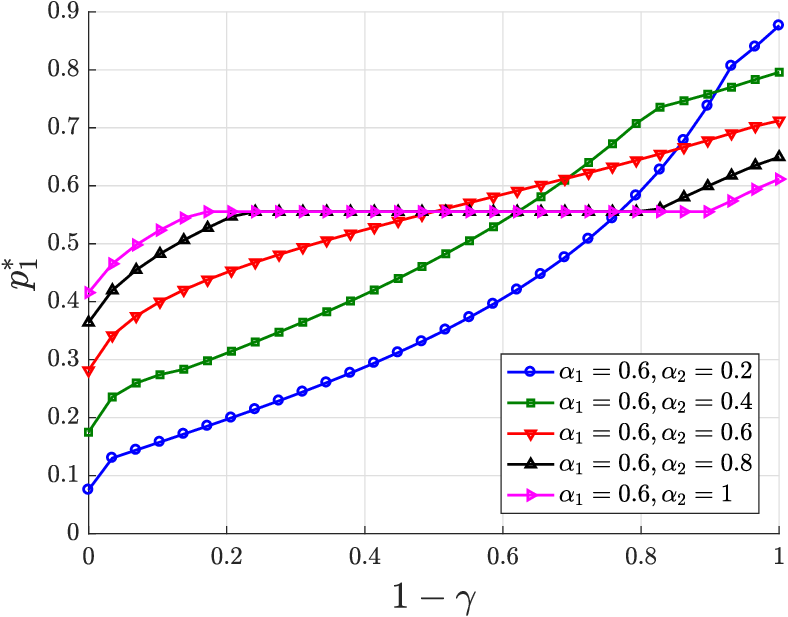}
\caption{Price $p_{1m}^*$ as a function of $1-\gamma$ for different values of $\alpha_1$ and  $\alpha_2$ (monopolistic)} \label{p1monhet}
\end{figure}

\begin{figure}%[!h]
\centering
\includegraphics[width=0.7\columnwidth]{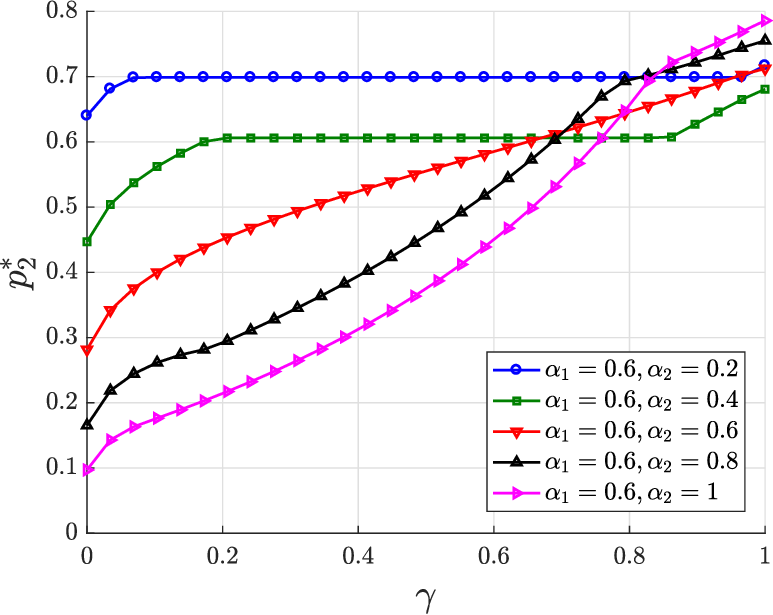}
\caption{Price $p_{2m}^*$ as a function of $\gamma$ for different values of $\alpha_1$ and $\alpha_2$ (monopolistic)} \label{p2monhet}
\end{figure}

Figures~\ref{p1monhet} and~\ref{p2monhet} show the prices $p_{1m}^*$ and $p_{2m}^*$ set by the NO, respectively, in the equilibrium to each subscriber base as a function of the service priority ($1-\gamma$ in Fig.~\ref{p1monhet} and $\gamma$ in Fig.~\ref{p2monhet}) for different values of $\alpha_2$.

It is seen that the service is priced higher as the service priority increases. And that the variation of the VO's subscribers sensitivity has an effect not only on the price set to the VO's subscribers (Fig.~\ref{p2monhet}), but also on the price set to the NO's subscribers (Fig.~\ref{p1monhet}). Indeed, increasing $\alpha_2$, while keeping $\alpha_1$ constant, causes $p_{1m}^*$ to increase (for low service priorities) or to decrease (for high service priorities). Additionally, a qualitative and relevant difference arises when compared with the case of common $\alpha$ (Figs.~\ref{p1monhom}): when $\alpha_1>\alpha_2$ (when $\alpha_1<\alpha_2$), the price set to the NO's subscribers, $p_{1m}^*$ (to the VO's subscribers, $p_{2m}^*$), does not vary for a range of intermediate values of $\gamma$.

\paragraph{Number of subscribers}

\begin{figure}%[!h]
\centering \includegraphics[width=0.7\columnwidth]{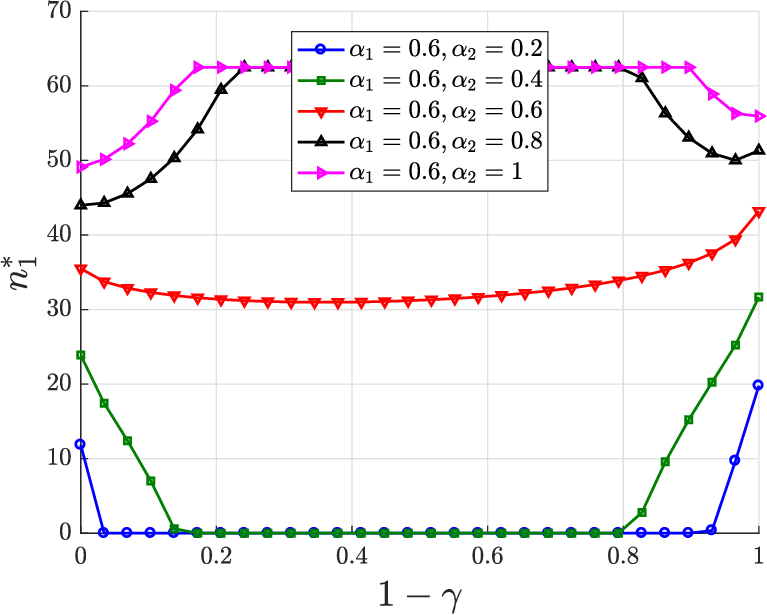}
\caption{Number of subscribers $n_{1m}^*$ as a function of $1-\gamma$ for different values of $\alpha_1$ and  $\alpha_2$ (monopolistic)} \label{n1monhet}
\vspace{-3mm}
\end{figure}

\begin{figure}%[!h]
\centering
\includegraphics[width=0.7\columnwidth]{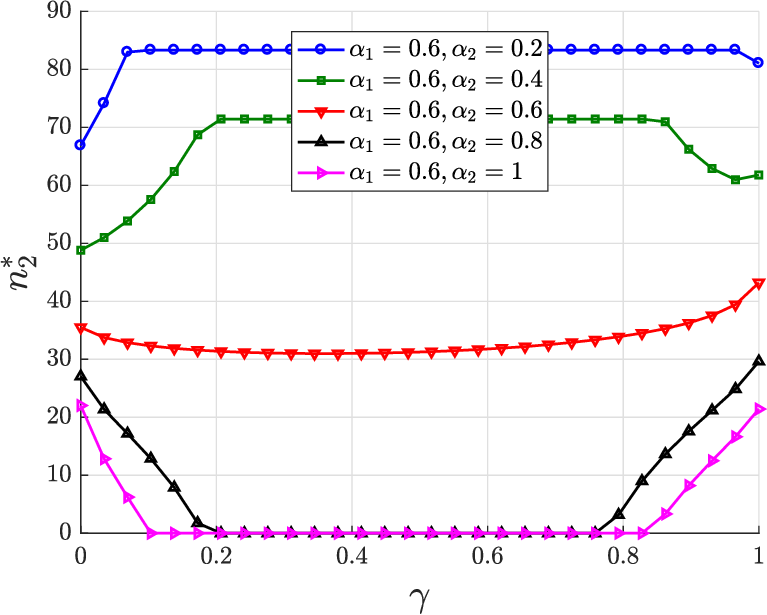}
\caption{Number of subscribers $n_{2m}^*$ as a function of $\gamma$ for different values of $\alpha_1$ and  $\alpha_2$ (monopolistic)} \label{n2monhet}
\end{figure}

Figures~\ref{n1monhet} and~\ref{n2monhet} show the number of subscribers from each user base, $n_{1m}^*$ and $n_{2m}^*$, respectively, in the equilibrium as a function of the service priority ($1-\gamma$ in Fig.~\ref{n1monhet} and $\gamma$ in Fig.~\ref{n2monhet}) for different values of $\alpha_2$.

It is seen that again the variation of the VO's subscribers sensitivity has an effect not only on the number of VO's subscribers (Fig.~\ref{n2monhet}), but also on the number of NO's subscribers (Fig.~\ref{n1monhet}). Indeed, increasing $\alpha_2$, while keeping $\alpha_1$ constant, causes the number of VO's subscribers to decrease and the number of NO's subscribers to increase. This is consistent with the fact that VO's subscribers become more sensitive to the delay.

Additionally, two qualitative and relevant differences arise when compared with the case of common $\alpha$ (Figs.~\ref{n1monhom}): when $\alpha_1>\alpha_2$ (when $\alpha_1<\alpha_2$), the number of NO's subscribers, $n_{1m}^*$ (VO's subscribers, $n_{2m}^*$), reaches a maximum and $n_{2m}^*$ ($n_{1m}^*$) goes down to zero for a range of intermediate values of $\gamma$, and keeps constant. This is due to the fact that, when $\alpha_1>\alpha_2$, the equilibrium is reached at the boundary between Region~I and Region~II (see Figs.~\ref{regions1} and~\ref{regions2}); and, when $\alpha_1<\alpha_2$, at the boundary between Region~I and Region~III.

\begin{figure}%[!h]
\centering
\includegraphics[width=0.7\columnwidth]{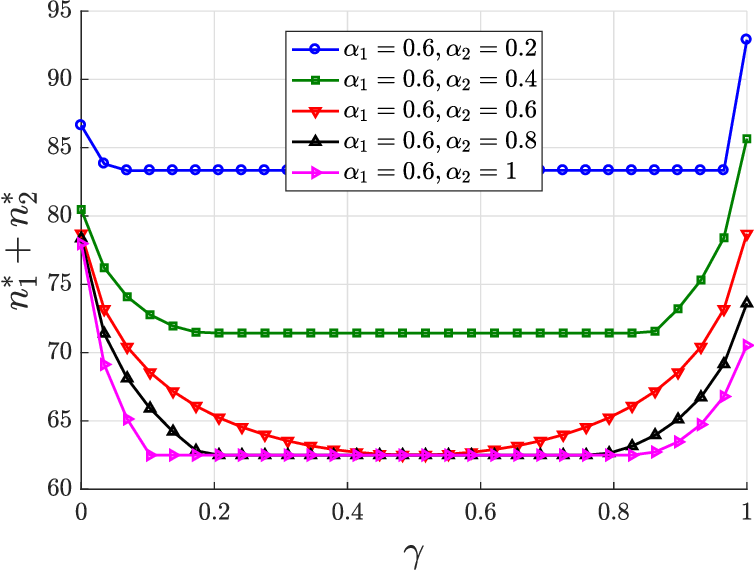}
\caption{Total number of subscribers as a function of $\gamma$ for different values of $\alpha_1$ and  $\alpha_2$ (monopolistic)} \label{n1n2monhet}
\end{figure}

Finally, Fig.~\ref{n1n2monhet} shows the total number of subscribers, i.e., $n_{1m}^*+n_{2m}^*$. Maximum subscription rates are achieved either when $\gamma$ is equal to 1 or to 0. Specifically, when $\alpha_1>\alpha_2$, the maximum subscription rate is achieved when $\gamma=1$ (or maximum service priority for VO's subscriber base). And vice versa when $\alpha_1<\alpha_2$.

The conclusion is that, when the two user bases have different sensitivities, the most favorable network configuration for the users is one where full priority is given to the lower sensitivity user base.

\paragraph{Operator's Profit}

\begin{figure}%[!h]
\centering
\includegraphics[width=0.7\columnwidth]{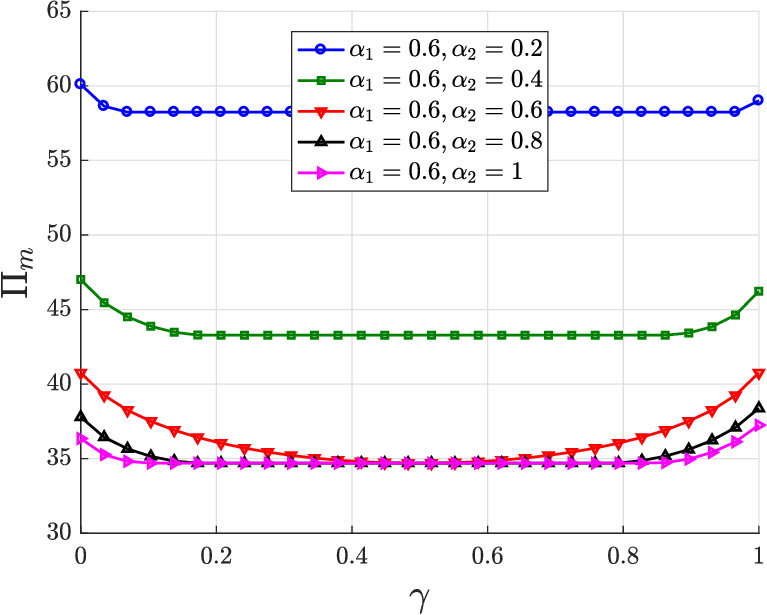}
\caption{NO's profits $\Pi_m^*$ as a function of $\gamma$ for different values of $\alpha_1$ and  $\alpha_2$ (monopolistic)} \label{pimonhet}
\end{figure}

Figure~\ref{pimonhet} shows NO's profit as a function of $\gamma$ (VO's subscriber priority) for different values of $\alpha_2$.

It is seen that similar (but no necessarily equal) priorities result in minimum profit. This is consistent with the fact that intermediate values for $\gamma$ result in either no NO's subscribers ($\alpha_1>\alpha_2$ in Fig.~\ref{n1monhet}) or no VO's subscribers ($\alpha_1<\alpha_2$ in Fig.~\ref{n2monhet}). The maximum profits are achieved when $\gamma$ is equal to 1 or to 0. Specifically, when $\alpha_1>\alpha_2$, the maximum profit is achieved when $\gamma=0$ (or maximum service priority for NO's subscriber base). And vice versa when $\alpha_1<\alpha_2$.

The conclusion is that, when the two user bases have different sensitivities, the most favorable network configuration for the operator is one where full priority is given to higher sensitivity user base. This is in contrast with the most favorable configuration from the point of view of the users.
 
\FloatBarrier

%--------------------------------------------------------------------------
\subsection{Strategic business model}\label{sec:results_strategic}
%--------------------------------------------------------------------------

The equilibrium in the strategic scenario, where the NO operates the network and provides service to its own subscriber base and the VO provides service to its own subscriber base, is presented and discussed. The NO's subscriber base is assigned a priority $1-\gamma$ in the use of the network capacity, while the VO's subscriber base is assigned a priority $\gamma$. Finally, the VO pays a fee $\delta$ to the NO for each VO's subscriber.

First, the case where both user bases have the same sensitivity to the delay, i.e, $\alpha_1=\alpha_2$, is addressed, and second, the case where they have different sensitivities, i.e., $\alpha_1\neq\alpha_2$. 

Two caveats should be remarked in the interpretation of the results discussed in this section. First, the strategy profile $\{ p_1^*, p_2^* \} =\{0,0\}$ is always one Nash equilibrium, but no explicit reference to it is made thereafter, since there is always a non-zero Nash equilibrium. 
And second, there are parameter configurations where there is a continuum of Nash equilibria. More specifically, there are Nash equilibria where $\underline{p_1} \leq p_1^* \leq \overline{p_1}$ and $ \underline{p_2} \leq p_2^* \leq \overline{p_2}$. In those cases, the chosen equilibrium is that where the aggregate profit $\Pi_1^*+\Pi_2^*$ is maximum.

%--------------------------------------------------------------------------
\subsubsection{User bases with the same sensitivity}

Parameters are set $\alpha_1=\alpha_2=0.6$ and the effect of the fee $\delta$ on the equilibrium is analyzed. \inserta{Values for $\delta$ have been chosen so that both the NO and the VO obtain positive profits.}\label{tti:R2c11c}

\paragraph{Prices} 

\begin{figure}%[!h]
\centering
\includegraphics[width=0.7\columnwidth]{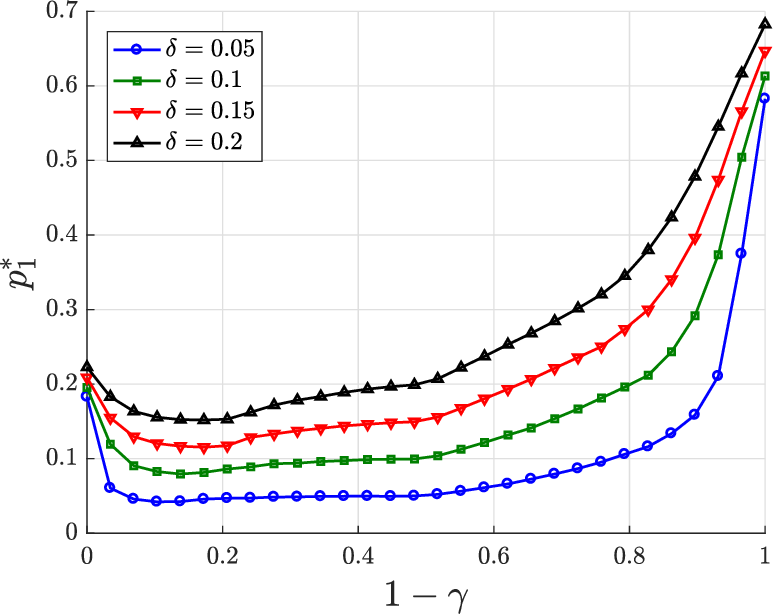}
\caption {Price $p_1^*$ as a function of $1-\gamma$ for different values of $\delta$ (strategic)} \label{p1sthom}%\label{p1completo}
\end{figure}

\begin{figure}%[!h]
\centering
\includegraphics[width=0.7\columnwidth]{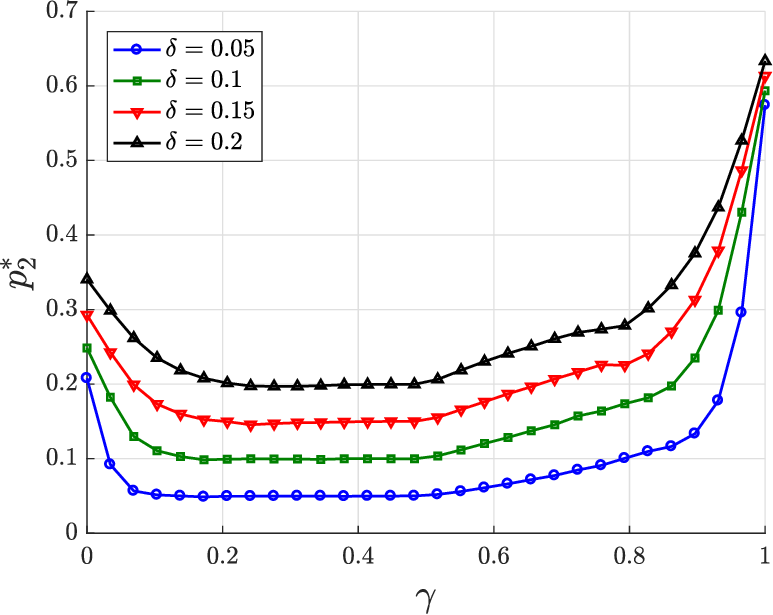} 
\caption {Price $p_2^*$ as a function of $\gamma$ for different values of $\delta$ (strategic)} \label{p2sthom}%\label{p2completo}
\end{figure}

Figures~\ref{p1sthom} and~\ref{p2sthom} show the prices $p_{1}^*$ set by the NO and $p_{2}^*$ set by the VO, respectively, in the equilibrium to their respective subscriber base as a function of the service priority ($1-\gamma$ in Fig.~\ref{p1sthom} and $\gamma$ in Fig.~\ref{p2sthom}) for different values of $\delta$.

It is seen that, for a range of priority values greater than a threshold, the service is priced higher as the priority increases. \inserta{The threshold for the NO's priority if different from the threshold for the VO's priority. For the NO, the threshold varies between 0.5 (when $\delta=0.05$) and~0.75 (when $\delta=0.2$). For the VO, the threshold is an interval of priorities where the price flattens and it varies between $[0.1,0.5]$ (when $\delta=0.05$) and~$[0.25,0.5]$ (when $\delta=0.2$)}\label{tti:R2c14}. And that, below that threshold, the service is priced higher as the priority decreases. This behavior is different from the behavior observed in the monopolistic scenario. The reason lies in the fact that a strategic interaction between the two operators takes place here and was absent there. It is also seen that the subscriber base that receives a service supported by a greater priority is priced higher; consistently with this observation, when $\gamma=1/2$, both subscriber bases are priced equally.

As regards the effect of $\delta$, it is seen that the higher $\delta$ is, the higher both prices are.

\paragraph{Number of subscribers}

\begin{figure}%[!h]
\centering
\includegraphics[width=0.7\columnwidth]{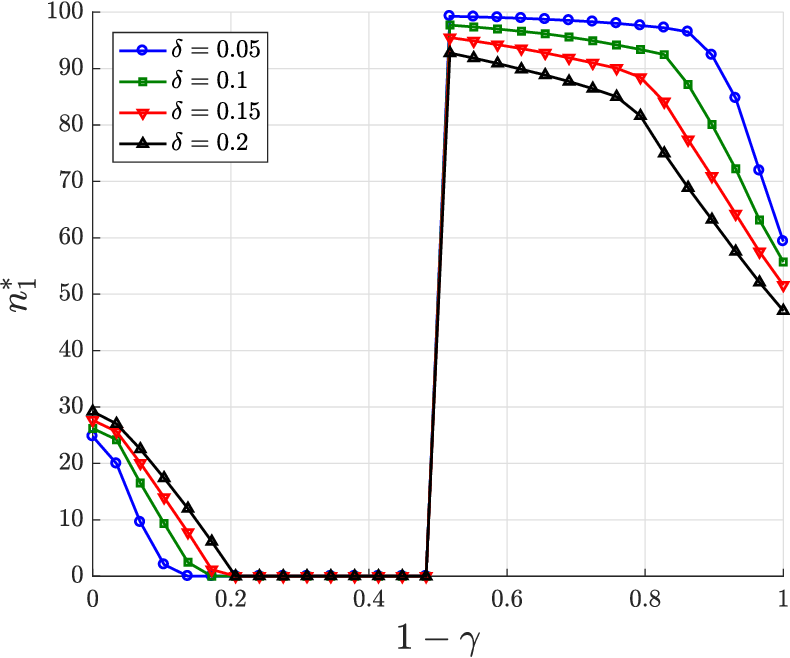} 
\caption {Number of subscribers $n_1^*$ as a function of $1-\gamma$ for different values of $\delta$ (strategic)} \label{n1sthom}%\label{n1Total}
\end{figure}

\begin{figure}%[!h]
\centering
\includegraphics[width=0.7\columnwidth]{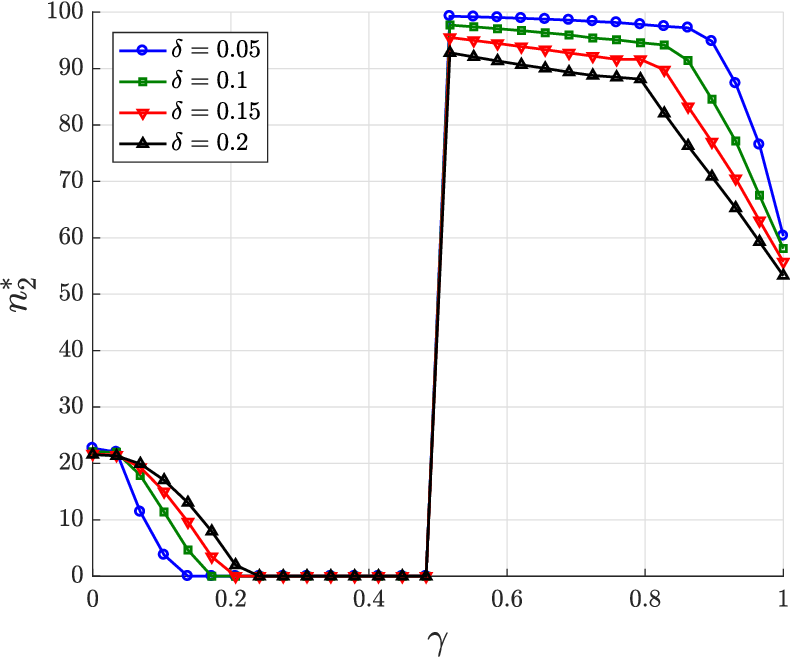}
\caption {Number of subscribers $n_2^*$ as a function of $\gamma$ for different values of $\delta$ (strategic)} \label{n2sthom}%\label{n2Total}
\end{figure}

Figures~\ref{n1sthom} and~\ref{n2sthom} show the number of NO's subscribers $n_{1}^*$ and of VO's subscribers $n_{2}^*$, respectively, in the equilibrium, as a function of the service priority ($1-\gamma$ in Fig.~\ref{n1sthom} and $\gamma$ in Fig.~\ref{n2sthom}) for different values of $\delta$.

It is seen that, for priority values greater than $1/2$, the number of subscribers decreases from a maximum value as the priority increases. This is consistent with the fact that the price (e.g., $p_1^*$ in Fig.~\ref{p1sthom}) increases, which seems to have a negative effect that is not compensated by the improvement of the QoS. The effect of an increase of $\delta$ is to increase $n_{1}^*$ and to decrease $n_{2}^*$, which is consistent with the increase in the fee per subscriber that the VO should pay.

On the other hand, for priority values lower than $1/2$, the number of subscribers  decreases as the priority increases until it collapses to zero beyond a threshold priority. The effect of $\delta$ is a mixed one, depending on the value of the priority. The absence of subscribers for a range of priorities lower than $1/2$ but not very low is caused by a combined effect of the relatively low price and high QoS for the other operator's user base and of the relatively low QoS for the own user base. Note that for this range of priorities values, the number of subscribers is maximum for the other operator's user base.

\begin{figure}%[!h]
\centering
\includegraphics[width=0.7\columnwidth]{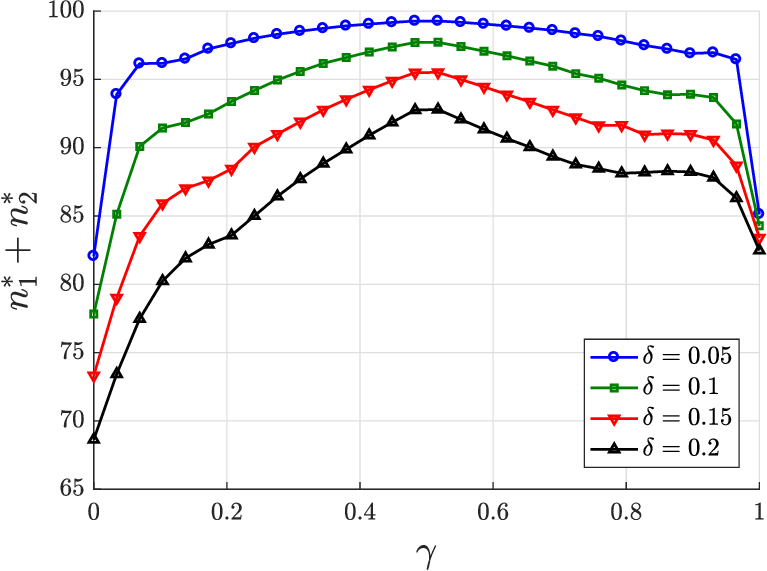}
\caption {Total number of subscribers as a function of $\gamma$ for different values of $\delta$ (strategic)} \label{n1n2sthom}
\end{figure}

Finally, Fig.~\ref{n1n2sthom} shows the total number of subscribers, i.e., $n_{1m}^*+n_{2m}^*$. Maximum subscription rates are achieved when $\gamma$ is equal to $1/2$, which is a different result from the monopolistic business model.

The conclusion is that, when the two user bases have the same sensitivity, the most favorable network configuration for the users is one where the priority is equally shared between the user bases. 

\paragraph{Profit}

\begin{figure}%[h]
\centering
\includegraphics[width=0.7\columnwidth]{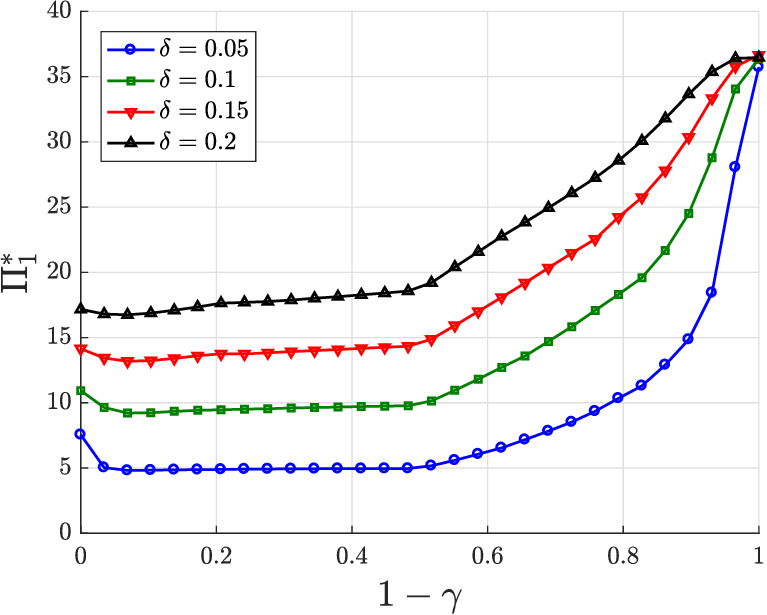}
\caption {NO's profit $\Pi_1^*$ as a function of $1-\gamma$ for different values of $\delta$ (strategic)} \label{pi1sthom}%\label{Pi1completo}
\end{figure}

\begin{figure}%[h]
\centering
\includegraphics[width=0.7\columnwidth]{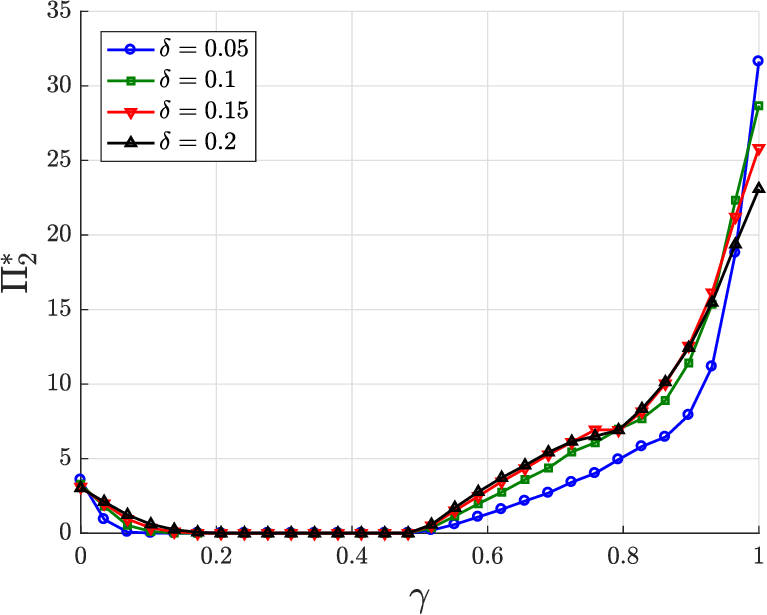}
\caption {VO's profit $\Pi_2^*$ as a function of $\gamma$ for different values of $\delta$ (strategic)} \label{pi2sthom}%\label{Pi2completo}
\end{figure}

Figures~\ref{pi1sthom} and~\ref{pi2sthom} show the NO's profit $\Pi_1^*$ and the VO's profit $\Pi_2^*$, respectively, in the equilibrium as a function of the service priority ($1-\gamma$ in Fig.~\ref{pi1sthom} and $\gamma$ in Fig.~\ref{pi2sthom}) for different values of $\delta$.

It is seen that the effect of the priority on the profit is similar to the effect on the price, that is, the higher the priority, the higher the profits, except for very low priorities. And also that the modulating effect that the number of subscribers has on the profits: when $n_2^*$ is zero in Fig.~\ref{n2sthom}, the profit $\Pi_2^*$ is zero in Fig.~\ref{pi2sthom}. However, when $n_1^*$ is zero in Fig.~\ref{n1sthom}, the profit $\Pi_1^*$ is not zero in Fig.~\ref{pi1sthom}, because the NO gets revenue also from the VO's subscriber base, which is not zero in Fig.~\ref{n2sthom}.

The effect of the fee $\delta$ paid by the VO to the NO on the profit $\Pi_1^*$ is clear in Fig.~\ref{pi1sthom}: the higher the $\delta$, the higher the $\Pi_1^*$. The effect on the profit $\Pi_2^*$ (Fig.~\ref{pi2sthom}) is not so clear: the higher the $\delta$, the lower the $\Pi_2^*$, only for high priorities.

The conclusion is that, for user bases with the same sensitivity, the most favorable network configuration for each operator is one where all priority is given to its own user base. This is in contradiction with the most favorable configuration from the point of view of the users.

\FloatBarrier

%--------------------------------------------------------------------------
\subsubsection{User bases with different sensitivities} 

Parameters are set as $\alpha_1=0.6$ and $\alpha_2=\{0.2, 0.4, 0.6, 0.8, 1\}$ and the effect of the priority $\gamma$ on the equilibrium is analyzed. Parameter $\delta$ is set to 0.15.

\paragraph{Prices}

\begin{figure}%[!h]
\centering
\includegraphics[width=0.7\columnwidth]{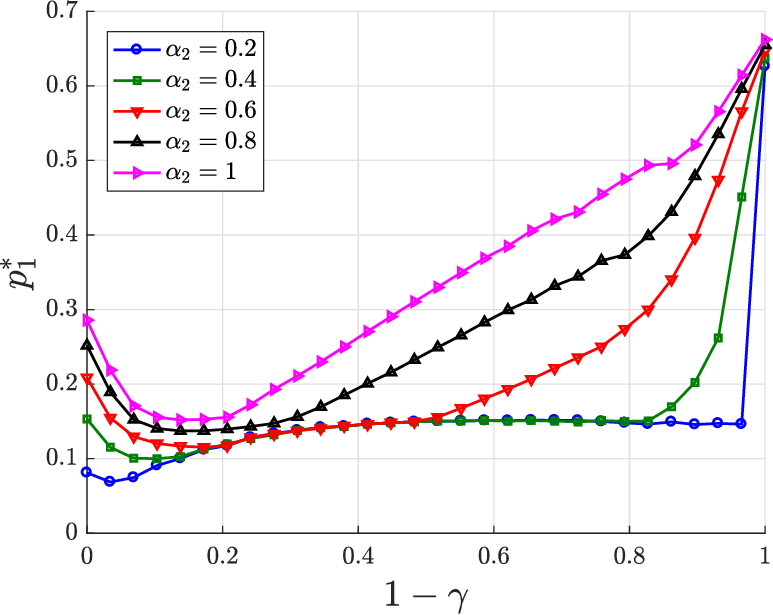}
\caption {Price $p_1^*$ as a function of $1-\gamma$ for different values of $\alpha_2$ (strategic scenario, $\alpha_1=0.6$, $\delta=0.15$)} \label{p1sthet}
\end{figure}

\begin{figure}%[!h]
\centering
\includegraphics[width=0.7\columnwidth]{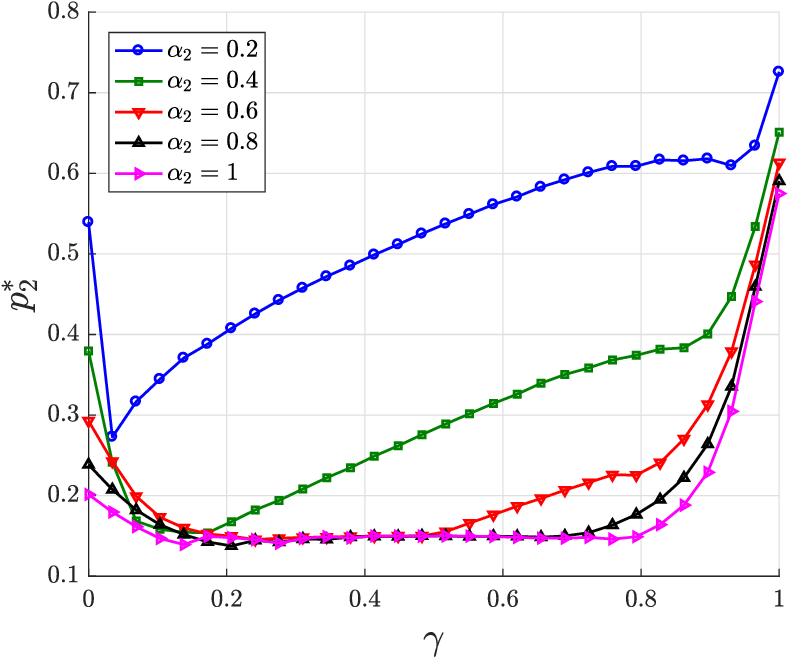} 
\caption {Price $p_2^*$ as a function of $\gamma$ for different values of $\alpha_2$ (strategic scenario, $\alpha_1=0.6$, $\delta=0.15$)} \label{p2sthet}%\label{p2completo}
\end{figure}

Figures~\ref{p1sthet} and~\ref{p2sthet} show the prices $p_{1}^*$ set by the NO and $p_{2}^*$ set by the VO, respectively, in the equilibrium as a function of the service priority ($1-\gamma$ in Fig.~\ref{p1sthet} and $\gamma$ in Fig.~\ref{p2sthet}) for different values of $\alpha_2$.

It is seen that there are similarities with the case of user bases with the same sensitivities as long as $\alpha_2 \geq 0.4$: both prices increase for priority values approaching 0 and 1. 
 
\paragraph{Number of subscribers}

\begin{figure}%[!h]
\centering
\includegraphics[width=0.7\columnwidth]{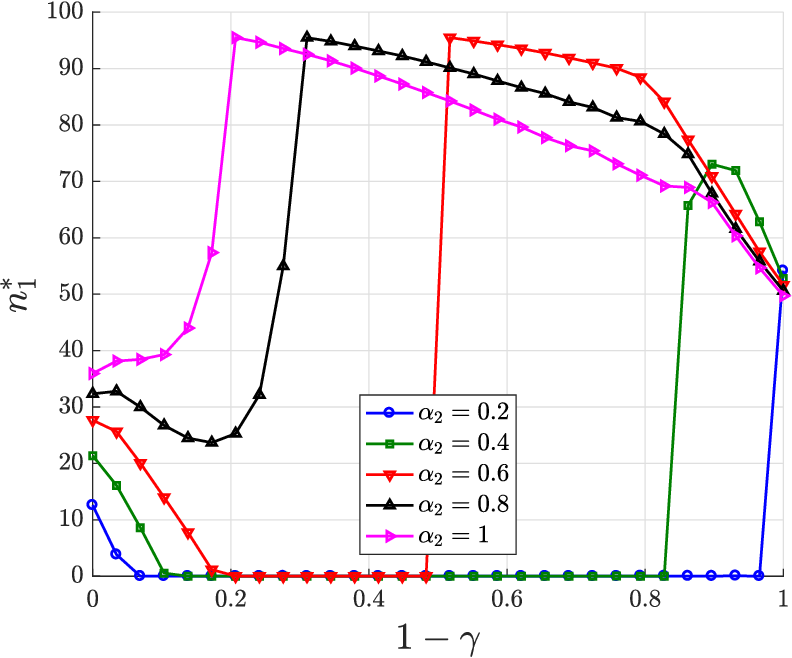} 
\caption {Number of subscribers $n_1^*$ as a function of $1-\gamma$ for different values of $\alpha_2$ (strategic scenario, $\alpha_1=0.6$, $\delta=0.15$)} \label{n1sthet}
\end{figure}

\begin{figure}%[!h]
\centering
\includegraphics[width=0.7\columnwidth]{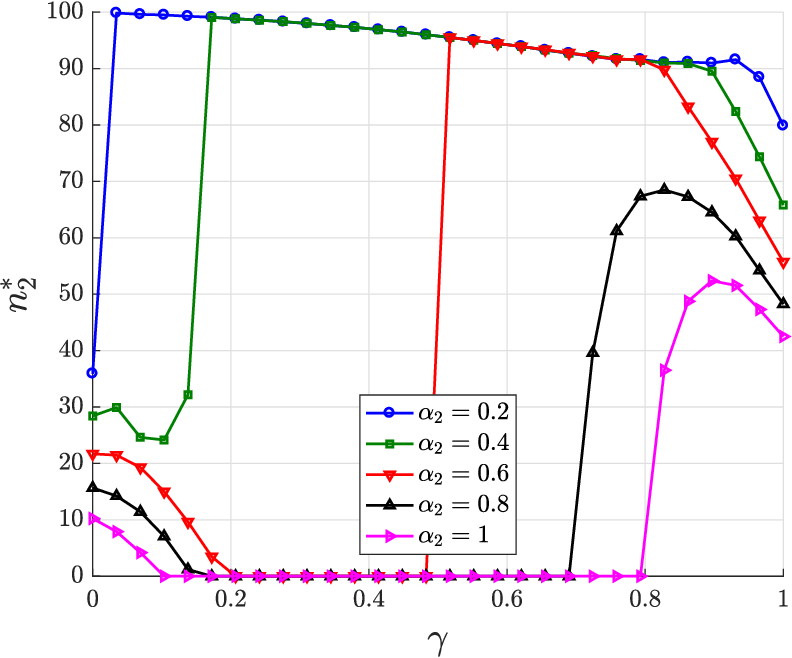}
\caption {Number of subscribers $n_2^*$ as a function of $\gamma$ for different values of $\alpha_2$ (strategic scenario, $\alpha_1=0.6$, $\delta=0.15$)} \label{n2sthet}
\end{figure}

Figures~\ref{n1sthet} and~\ref{n2sthet} show the number of NO's subscribers $n_{1}^*$ and of VO's subscribers $n_{2}^*$, respectively, in the equilibrium as a function of the service priority ($1-\gamma$ in Fig.~\ref{n1sthet} and $\gamma$ in Fig.~\ref{n2sthet}) for different values of $\alpha_2$.

It is seen that there is a sharp transition from a minimum to a maximum value of $n_1^*$, as in the case when the user bases had the same sensitivities. However, the minimum value is not always zero. And the priority value where the transition takes place depends on $\alpha_2$: the higher the $\alpha_2$, the higher the transition priority. 

\begin{figure}%[!h]
\centering
\includegraphics[width=0.7\columnwidth]{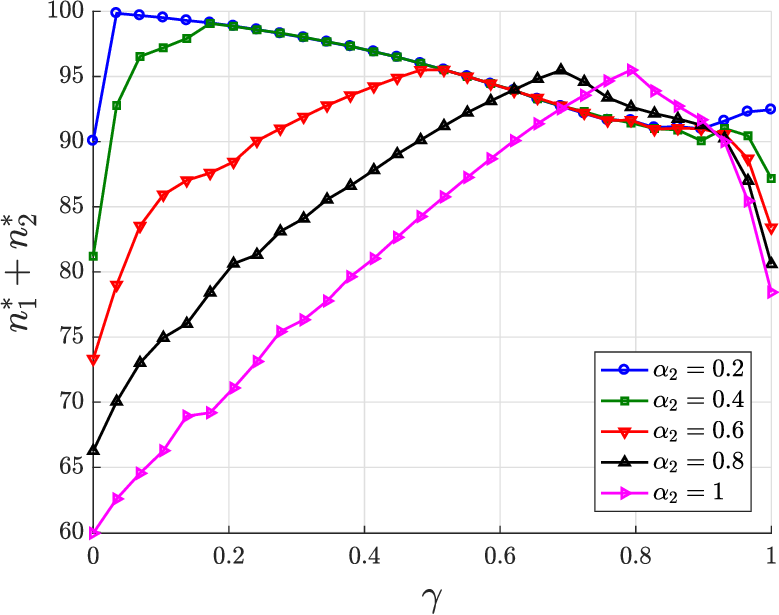}
\caption {Total number of subscribers as a function of $\gamma$ for different values of $\alpha_2$ (strategic scenario, $\alpha_1=0.6$, $\delta=0.15$)} \label{n1n2sthet}
\end{figure}

Finally, Fig.~\ref{n1n2sthet} shows the total number of subscribers, i.e., $n_{1m}^*+n_{2m}^*$. Maximum subscription rates are achieved for an intermediate value of $\gamma$, which is $1/2$ for the case with the same sensitivities. This intermediate value depends, however, on $\alpha_2$: the higher $\alpha_2$, the higher this intermediate value that results in a maximum subscription rate. 

The conclusion is that, when the two user bases have different sensitivities, the most favorable network configuration for the users is one where the priority is shared between the user bases, at a ratio that depends on how different the sensitivities are. 

\paragraph{Profit}

\begin{figure}%[h]
\centering
\includegraphics[width=0.7\columnwidth]{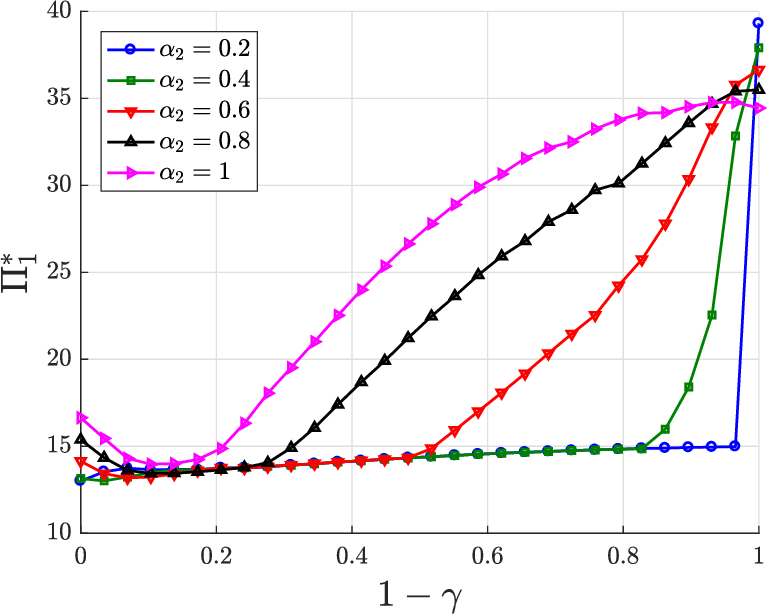}
\caption {NO's profit $\Pi_1^*$ as a function of $1-\gamma$ for different values of $\delta$ (strategic)} \label{pi1sthet}
\end{figure}

\begin{figure}%[h]
\centering
\includegraphics[width=0.7\columnwidth]{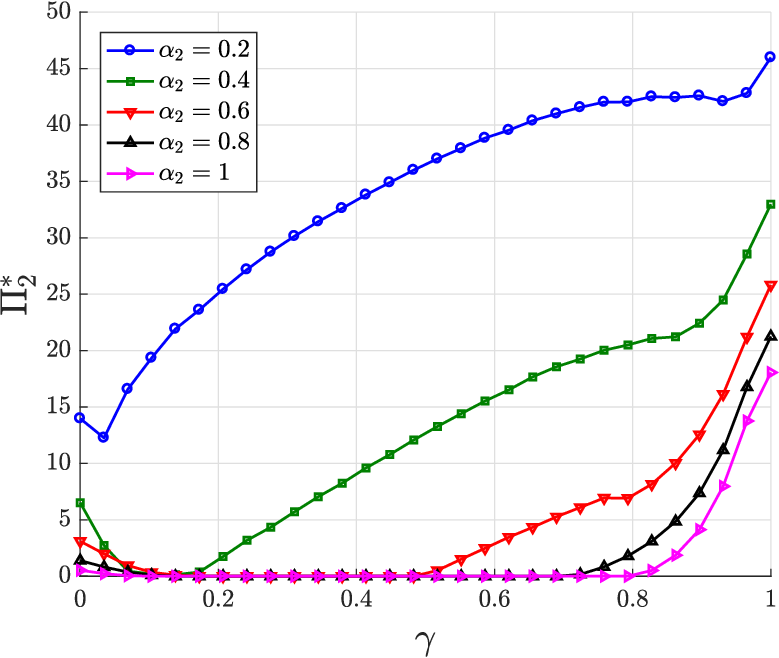}
\caption {VO's profit $\Pi_2^*$ as a function of $\gamma$ for different values of $\delta$ (strategic)} \label{pi2sthet}
\end{figure}

Figures~\ref{pi1sthet} and~\ref{pi2sthet} show the NO's profit $\Pi_1^*$ and the VO's profit $\Pi_2^*$, respectively, in the equilibrium as a function of the service priority ($1-\gamma$ in Fig.~\ref{pi1sthet} and $\gamma$ in Fig.~\ref{pi2sthet}) for different values of $\alpha_2$.

It is seen that the effect of the priority on the profit is similar to that of the case with the same sensitivities, that is, the higher the priority, the higher the profits, except for very low priorities (and except for the NO, $\alpha2=1$ and high priorities).

In addition, as the sensitivity of the VO's user base comparatively increases, the VO's profit decreases, while the NO's profit increases.

The conclusion is that, for user bases with different sensitivities, the most favorable network configuration for each operator is one where all priority is given to its own user base, which was the same conclusion reached when user bases had the same sensitivities. This conclusion is in contradiction with the most favorable configuration from the point of view of the users.

\FloatBarrier

%--------------------------------------------------------------------------
\subsection{Business model feasibility}\label{sec:results_feasibility}
%--------------------------------------------------------------------------

In this section, the conditions are discussed under which the business models proposed in Section~\ref{sec:model} are feasible, i.e., provide incentives to both the NO and the VO. First, the monopolistic business model is tackled, and then, the strategic one.

%--------------------------------------------------------------------------
\subsubsection{Monopolistic business model}

The conditions are searched under which the NO has an incentive to serve the VO's subscriber base and the VO has an incentive to let the NO do it.

The NO will acquiesce in providing service to the VO's subscriber base if the profit in the monopolistic scenario, $\Pi_m^*$, is greater than or equal to the profit in the baseline scenario, $\Pi_0^*$:
\begin{equation}\label{feasiblem1a}
\Pi_m^* \geq \Pi_0^*.
\end{equation}
The VO will agree to let the NO provide service to its subscriber base provided that it is not worse off than in the baseline scenario. Since the VO does not provide service nor gets any revenue in both the baseline and monopolistic scenario, this condition is always met.

A more realistic analysis would ask for a lump-sum payment $m > 0$ to be paid by the VO to the NO in order that the latter accepts the agreement. In this case, condition~\eqref{feasiblem1a} would transform into
\begin{equation}\label{feasiblem1b}
\Pi_m^* + m \geq \Pi_0^*,
\end{equation}
which is feasible if
\begin{equation}\label{feasiblem1c}
\Pi_m^* > \Pi_0^*.
\end{equation}

Figures~\ref{Profits04}, \ref{Profits06} and~\ref{Profits08} show $\Pi_m^*$ as a function of $\gamma$  when the VO's user base has sensitivity, $\alpha_2$, equal to 0.4, 0.6 and 0.8, respectively. In the three figures the NO's user base has sensitivity $\alpha_1=0.6$. The value $\Pi_0^*$ is also represented as a single (is independent of $\alpha_2$) horizontal (is independent of $\gamma$) line. The profit for the strategic business model is also represented, but it will be discussed in next subsection.
Condition~\eqref{feasiblem1c} is met for every $\gamma$ and $\alpha_2$. Nevertheless, the incentive is more scarce as $\alpha_2$ increases and as $\gamma$ approaches $1/2$, that is, as the VO's user base has a higher sensitivity and as both subscriber bases are serviced under more similar priorities.

%--------------------------------------------------------------------------
\subsubsection{Strategic business model}

Now the conditions are searched under which the NO has an incentive to acquiesce in the VO's entry and the VO has an incentive to enter into the infrastructure sharing agreement.

Again, the NO will acquiesce in the VO's entry if the profit in the strategic scenario, $\Pi_1^*$, is greater than or equal to the profit in the baseline scenario, $\Pi_0^*$:
\begin{equation}\label{feasibles1a}
\Pi_1^* \geq \Pi_0^*.
\end{equation}

The VO will enter the market and agree to pay for sharing the infrastructure if the profit in the strategic scenario, $\Pi_2^*$, is greater than or equal to the profit if it does not enter, which is assumed to be zero:
\begin{equation}\label{feasibles2a}
\Pi_2^* \geq 0.
\end{equation}

If~\eqref{feasibles1a} is not met, a lump-sum payment $m > 0$ could be agreed to be paid by the VO to the NO (apart from the fee per subscriber, $\delta$) for the latter to acquiesce in the entry, i.e.,
\begin{equation}\label{feasibles1b}
\Pi_1^*+m \geq \Pi_0^*.
\end{equation}
But now, condition~\eqref{feasibles2a} should be transformed into
\begin{equation}\label{feasibles2b}
\Pi_2^* -m\geq 0.
\end{equation}
From~\eqref{feasibles1b} and~\eqref{feasibles2b}, such a lump-sum payment $m$ can be guaranteed if 
\begin{equation}
\Pi_1^*+\Pi_2^* > \Pi_0^* \label{feasibles3}.
\end{equation}

\begin{figure}%[h!]
\centering
\includegraphics[width=0.6\textwidth]{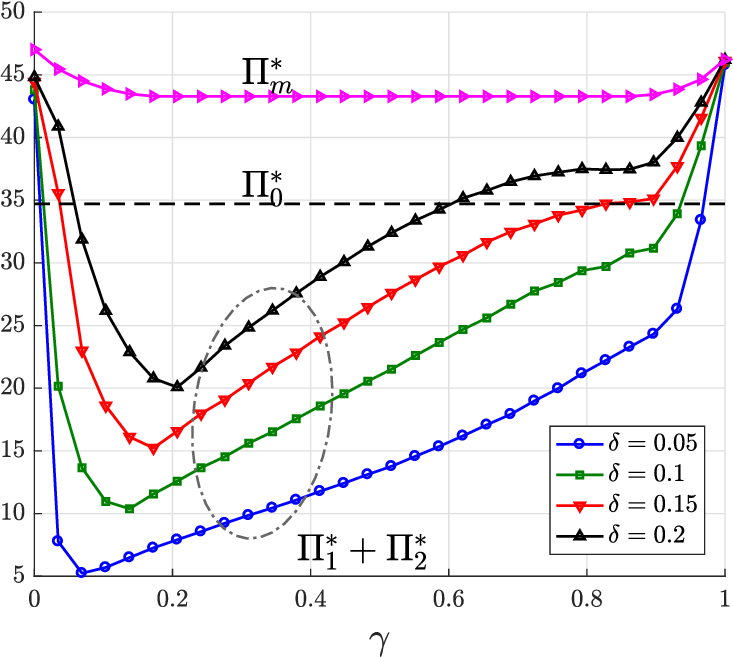}
\caption {$\Pi_1^*+\Pi_2^*$ and $\Pi_m^*$ as a function of $\gamma$ for $\alpha_1=0.6$ and $\alpha_2=0.4$ and different values of $\delta$} \label{Profits04} %\label{p1p22tt}
\end{figure}

\begin{figure}%[h!]
\centering
\includegraphics[width=0.6\textwidth]{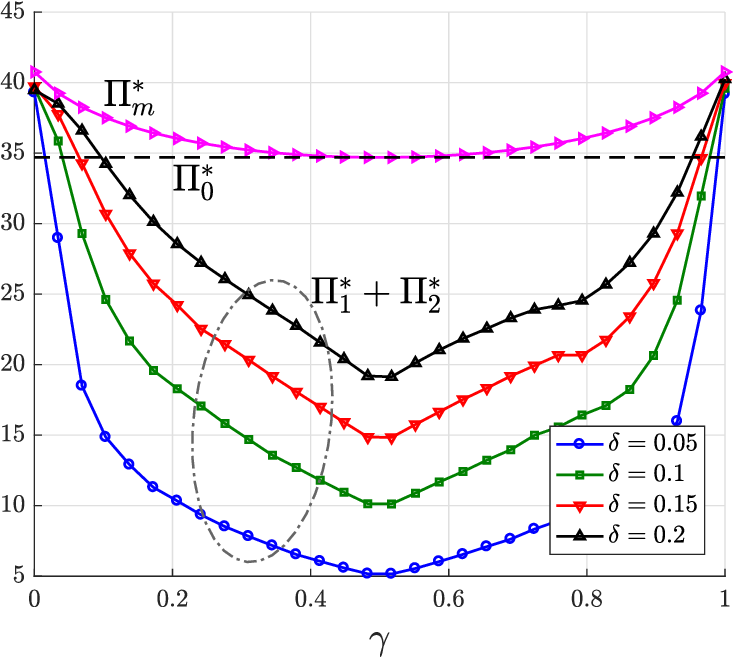}
\caption {$\Pi_1^*+\Pi_2^*$ and $\Pi_m^*$ as a function of $\gamma$ for $\alpha_1=0.6$ and $\alpha_2=0.6$ and different values of $\delta$} \label{Profits06} %\label{p1p22tt}
\end{figure}

\begin{figure}%[h!]
\centering
\includegraphics[width=0.6\textwidth]{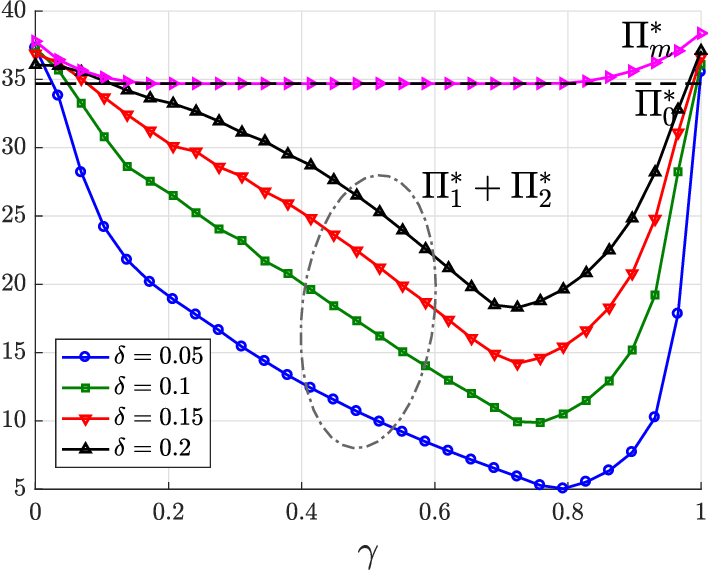}
\caption {$\Pi_1^*+\Pi_2^*$ and $\Pi_m^*$ as a function of $\gamma$ for $\alpha_1=0.6$ and $\alpha_2=0.8$ and different values of $\delta$} \label{Profits08} %\label{p1p22tt}
\end{figure}

Figures~\ref{Profits04}, \ref{Profits06} and~\ref{Profits08} also show $\Pi_1^*+\Pi_2^*$ for different values of $\delta$. Condition~\eqref{feasibles3} is only seen to be met for some values of the parameters $\alpha_2$, $\delta$ and $\gamma$. Specifically, 
\begin{itemize}
\item Given $\alpha_2$ and $\delta$, Condition~\eqref{feasibles3} is met for a range of $\gamma \in [0, \gamma_\text{low}] \cup [\gamma_\text{high}, 1 ]$, that is, for sufficiently different priorities for the two user bases.
\item For some $\alpha_2$ and $\gamma$, Condition~\eqref{feasibles3} is met for a sufficiently high $\delta$, that is, for a sufficiently high fee paid by the VO to the NO.
\item For some $\delta$ and $\gamma$, Condition~\eqref{feasibles3} is met for a sufficiently low $\alpha_2$, that is, for sufficiently low sensitivity VO's users.
\end{itemize}

The conclusion is that the two business models are feasible. The monopolistic business model for every parameter setting and the strategic one for the following situation: different priorities, not high sensitivity entrant users and high values of fee. 

The two business models are compared as follows. From the point of view of the profits, Figs.~\ref{Profits04}--\ref{Profits08} show that the monopolistic business model provides a stronger incentive to both the NO and the VO than the strategic one, since $\Pi_1+\Pi_2 \leq \Pi_m$ for every parameter setting. This results is obvious, because a strategic game always results in an outcome that is less than or equal the optimum outcome.

\begin{figure}%[h!]
\centering
\includegraphics[width=0.6\textwidth]{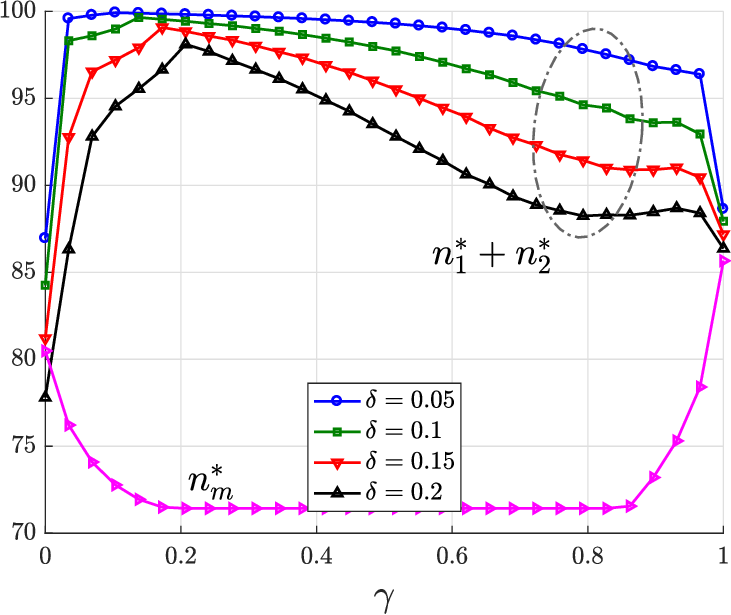}
\caption {$n_1^*+n_2^*$ in the strategic and the monopolistic scenarios as a fuction of $\gamma$ for $\alpha_1=0.6$ and $\alpha_2=0.4$ and different values of $\delta$} \label{Users04} %\label{p1p22tt}
\end{figure}

\begin{figure}%[h!]
\centering
\includegraphics[width=0.6\textwidth]{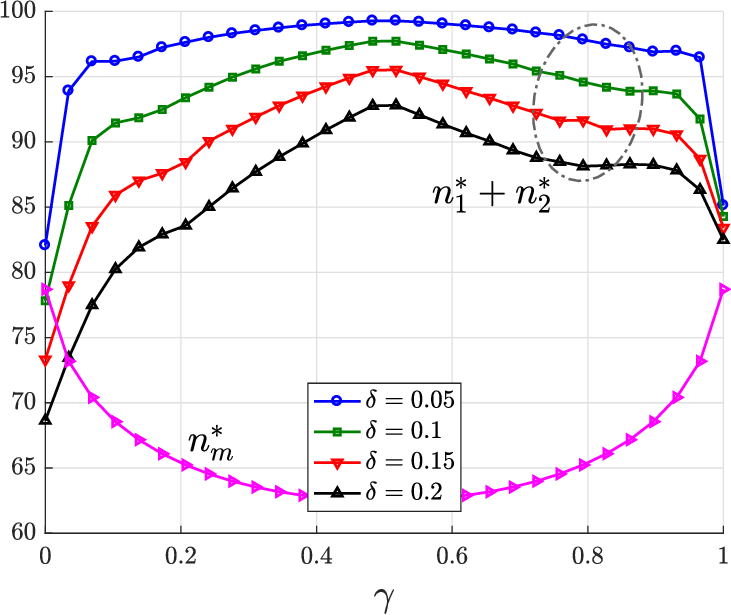}
\caption {$n_1^*+n_2^*$ in the strategic and the monopolistic scenarios as a fuction of $\gamma$ for $\alpha_1=0.6$ and $\alpha_2=0.6$ and different values of $\delta$} \label{Users06} %\label{p1p22tt}
\end{figure}

\begin{figure}%[h!]
\centering
\includegraphics[width=0.6\textwidth]{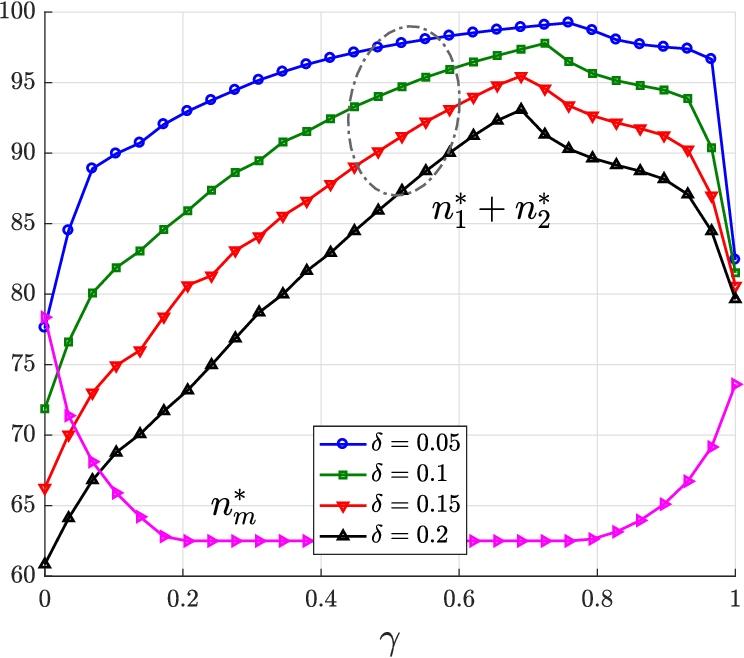}
\caption {$n_1^*+n_2^*$ in the strategic and the monopolistic scenarios as a fuction of $\gamma$ for $\alpha_1=0.6$ and $\alpha_2=0.8$ and different values of $\delta$} \label{Users08} %\label{p1p22tt}
\end{figure}

The business models can also be compared from the point of view of the users. More specifically, in terms of the number of subscribers that each business models supports. This indicator is a measure of the efficiency of the service provision. 
Figures~\ref{Users04}, \ref{Users06} and~\ref{Users08} show $n_1^*+n_2^*$ in the strategic (for different values of $\delta$) and the monopolistic business models as a function of $\gamma$. The NO's user base has sensitivity $\alpha_1=0.6$ in the three figures and the VO's user base has sensitivities $\alpha_2$ equal to 0.4 (Fig.~\ref{Users04}), 0.6 (Fig.~\ref{Users06}) and 0.8 (Fig.~\ref{Users08}). 
It is seen that the total number of subscribers in the strategic business model is greater than in the monopolistic one, for $\gamma$ higher that a threshold value that depends of $\alpha_2$ and $\delta$. For a given $\alpha_2$, the higher the $\delta$, the higher the threshold. And for a given $\delta$, the higher the $\alpha_2$,  the higher the threshold.

It can then be stated that the strategic business model is more desirable from the point of view of the users when the entrant user base has not a high sensitivity and/or it is serviced under a sufficiently high priority and the fee paid by the VO is not very high. Or, in other words, under this setting, the system is more efficient in the strategic business model.

%--------------------------------------------------------------------------
%--------------------------------------------------------------------------
\section{Conclusions and future work}\label{sec:conclusions}
%--------------------------------------------------------------------------
%--------------------------------------------------------------------------

Two business models have been proposed for providing service to two user bases over a common network infrastructure. The network resources are assigned to the user bases based on a priority sharing agreement, which can be supported by network slicing. In one business model, named \emph{monopolistic}, the network operator provides service to both user bases. In the other business model, named \emph{strategic}, the network operator provides service to its user base and a virtual operator provides service to its user base and pays a per-subscriber fee to the network operator.

It has been shown that both business models are feasible, since they provide incentives to both operators compared with the baseline scenario, where only the NO's user base receives service. The monopolistic business model provides a stronger incentive to the operators than the strategic business model, but the latter is more desirable from the point of view of the users, since a higher number of users receive service. 

This conclusion would provide a rationale for a regulator to enforce the entry of virtual operators in the market where there is an incumbent network operator.

\label{tti:R1c4}\inserta{As regards the future lines of work, three open issues are worth mentioning.
First, while keeping the M/M/1-DPS queue as a model for a sliced network, new business models could be modeled and analyzed. For example, the network could be operated by an wholesale network operator, that is, an operator not providing service to end users but supplying capacity to a set of virtual operators. The focus of the analysis would be then the capacity market between the NO and the VOs.
Second, the analysis conducted in this paper has comprised two scenarios: one where the users bases had the same sensitivities and one where they had different sensitivities. The utility expression for all users has been the same. An improved appraisal of the user heterogeneity could be obtained by modeling each user base with a different utility expression.
And thirdly, an issue that receives frequent attention from the literature is the investment incentives. Indeed, the conducted analysis in this paper could be extended, if expressions for the equilibrium could be simplified, by incorporating an investment decision to be made by the NO, who will anticipate the equilibrium of the two-stage pricing game for each business model.}

%-------------------------------------------------------------------------------------------------
%-------------------------------------------------------------------------------------------------
\section*{Declarations of interest, acknowledgement, funding sources and author contributions}
%-------------------------------------------------------------------------------------------------
%-------------------------------------------------------------------------------------------------

Declarations of interest: none.

The authors would like to thank the collaboration of Angel Sanchis-Cano in the early stages of this work.

This work has been supported by the Spanish Ministry of Science, Innovation and Universities (MCIU/AEI) and the European Union (FEDER/UE) through Grant PGC2018-094151-B-I00 and partially supported by the Salesian Polytechnic University of Ecuador through a Ph.D. scholarship granted to the first author.

E.~Sacoto contributed to the formal analysis, the methodology, the investigation and the writing of the original draft.
L.~Guijarro contributed to the conceptualization, the investigation, the review \& editing, the supervision and the funding acquisition.
J.R.~Vidal contributed to the investigation, the data curation, the review \& editing, and the visualization.
V.~Pla contributed to the formal analysis, the review \& editing, and the funding acquisition.

%\clearpage

%-------------------------------------------------------------------------------------------------
%-------------------------------------------------------------------------------------------------
%\section*{References}
%-------------------------------------------------------------------------------------------------
%-------------------------------------------------------------------------------------------------

\bibliographystyle{elsarticle-num}

\end{document}